\documentclass[fleqn,usenatbib,useAMS]{mnras}

\usepackage{newtxtext,newtxmath}

\usepackage[T1]{fontenc}
\usepackage{ae,aecompl}

\usepackage{graphicx}	
\usepackage{url}

\hypersetup{draft}

\def\[{\begin{equation}}
\def\]{\end{equation}}
\def\citejap#1{\citeauthor{#1}\ \citeyear{#1}}
\def\eqref#1{(\ref{#1})}

\def\hompc{\,h\,{\rm Mpc}^{-1}}

\newcommand{\WIxSC}{WI$\times$SC}
\newcommand{\phz}{photo-$z$}
\newcommand{\phzs}{photo-$z$s}
\newcommand{\zph}{z_\mathrm{phot}}
\newcommand{\zsp}{z_\mathrm{spec}}
\newcommand{\de}{\mathrm{d}}

\def\gs{\mathrel{\lower0.6ex\hbox{$\buildrel {\textstyle >}
 \over {\scriptstyle \sim}$}}}
\def\ls{\mathrel{\lower0.6ex\hbox{$\buildrel {\textstyle <}
 \over {\scriptstyle \sim}$}}}

\title[Wide-area tomography of CMB lensing]
{Wide-area tomography of CMB lensing and the growth of cosmological density fluctuations}

\author[J.A. Peacock and M. Bilicki]
{J.A. Peacock$^{1}$\thanks{E-mail: jap@roe.ac.uk}
and M. Bilicki$^{2,3,4}$\thanks{E-mail: bilicki@strw.leidenuniv.nl}
\\
$^{1}$Institute for Astronomy, University of Edinburgh, Royal Observatory, Edinburgh EH9 3HJ, United Kingdom \\
$^{2}$Leiden Observatory, Leiden University, Niels Bohrweg 2, NL-2333 CA Leiden, the Netherlands \\
$^{3}$National Centre for Nuclear Research, Astrophysics Division, P.O. Box 447, PL-90-950 {\L}\'{o}d\'{z}, Poland \\
$^{4}$Janusz Gil Institute of Astronomy, University of Zielona G\'ora, ul. Szafrana 2, 65-516 Zielona G\'{o}ra, Poland 
}

\pubyear{2018}

\begin{document}
\label{firstpage}
\pagerange{\pageref{firstpage}--\pageref{lastpage}}
\maketitle

\begin{abstract}
We describe a tomographic dissection of the {\it Planck\/} CMB lensing data,
cross-correlating this map with galaxies in different ranges of
photometric redshift.  We use the nearly all-sky 2MPZ and
WISE$\times$SCOS catalogues for $z<0.35$, extending to $z<0.6$ using
SDSS.  We describe checks for consistency between the different
datasets, and perform a test for possible leakage of thermal
Sunyaev--Zel'dovich signal into our cross-correlation measurements.
The amplitude of the cross-correlation allows us to estimate the
evolution of density fluctuations as a function of redshift, thus
providing a test of theories of modified gravity. Assuming the common
parametrisation for the logarithmic growth rate,
$f_g=\Omega_m(z)^\gamma$, we infer $\gamma=0.77\pm 0.18$ when
$\Omega_m$ is fixed using external data. Thus CMB lensing tomography
is currently consistent with Einstein gravity, where $\gamma=0.55$ is
expected.  We discuss how such constraints may be expected to improve
with future data.
\end{abstract}




\section{Introduction}

One of the more remarkable results of the {\it Planck\/} mission has been
its measurement of the impact of foreground mass inhomogeneities on
cosmic microwave background (CMB) fluctuations. Gravitational lensing by large-scale structure (LSS)
distorts the background Gaussian CMB sky and imprints non-Gaussian signatures,
whose detection allows the degree of gravitational lensing to
be inferred; this in turn yields a map of the projection of 
density fluctuations times a distance-dependent kernel (see e.g \citealt{LewisChallinor2006}). We then
obtain an astonishing picture containing imprints of every void or supercluster that
ever existed, projected against the backlight of the CMB.

CMB lensing was first detected by
\cite{Smith07} from cross-correlation of the Wilkinson Microwave
Anisotropy Probe (WMAP) data with radio galaxy counts from the NRAO
VLA Sky Survey (NVSS), and then confirmed by \cite{Hirata08} where
WMAP with a more extended set of LSS tracers was used. The first
measurements of this effect directly from auto-correlation of CMB data
were presented by \cite{Das11} using the Atacama Cosmology Telescope
(ACT) and by \cite{vanEngelen12} from the South Pole Telescope (SPT). 
\cite{Holder2013} also showed that the SPT lensing map
correlated with the high-$z$ cosmic infrared background.
These results were followed by all-sky CMB lensing analyses from the
{\it Planck\/} satellite via reconstruction techniques.  The initial
implementation of this approach in \cite{Planck_lens2013} was
then improved by including polarization data, which raised the total
$S/N$ of lensing detection to about 40 (\citejap{Planck_lens2015}).

The main effect derives from mass fluctuations at redshifts $z\simeq 2$,
set by a balance between geometrical factors that favour distant
lenses, and the fact that LSS grows with time.
This measure of structure at intermediate redshifts is a valuable complement
to the intrinsic CMB fluctuations at $z\gg1$, and breaks degeneracies that
exist between cosmological parameters inferred using CMB data
alone \citep{Sherwin11}. The {\it Planck\/} lensing measurements are closely consistent with
a standard flat $\Lambda$CDM model with $\Omega_m\simeq 0.3$,
and provide strong evidence for this model independent of alternative
powerful probes (SNe; BAO); see \cite{Planck_pars2015},
\cite{SN_JLA}, \cite{Alam2017}.

But although the CMB lensing kernel peaks at high redshift, the signal
is broadly distributed and significant lensing contributions
are made from LSS down to local redshifts, $z\sim0.1$
(\citejap{LewisChallinor2006}). This opens the possibility of
using CMB lensing to measure the growth of structure with time, provided
the contributions to CMB lensing from the different redshifts can be 
disentangled -- which can be achieved via cross-correlation
if we have a set of foreground galaxies of known redshift. In that
case, one can carry out a tomographic analysis in which the galaxies
are split into a number of broad redshift bins; for each of these
the galaxy autocorrelation and cross-correlation with the CMB
lensing signal can be measured. Both these correlations are proportional
on large scales to the matter power spectrum at the redshift concerned, times either 
$b^2$ or $b$ for auto- and cross-correlation, where $b$ is the
linear galaxy bias parameter. Thus both $b$ and the amplitude of
matter fluctuations as a function of redshift can be inferred.

Such measurements of the growth of density fluctuations are of great
interest. At the simplest level, the linear growth history is
predicted once the cosmological parameters are set; thus growth
measurements are useful additional information helping to pin down the parameters.
But the real interest comes in looking for non-standard outcomes,
particularly as a probe of the correct theory of gravity. Motivation
for studying non-standard gravity comes in turn from the late-time accelerated
cosmic expansion: the speculation is that this may reflect deviations in
the strength of gravity that become important at low redshifts,
thus altering the rate at which structure develops.
A comprehensive survey of possible models of modified gravity is
given by \cite{Clifton12}, although it should be noted that
the recent demonstration that gravitational waves travel at the
speed of light has had a major impact on the landscape of
possibilities (e.g. \citejap{Baker17}). But in any case, it
is common to approach the issue of structure
growth in an empirical manner
through the following simple parametrised form:
\[
{d\ln\delta\over d\ln a} = \Omega_m(a)^\gamma
\]
\citep{Linder05}, where $\delta$ is matter overdensity and $a(t)$ is the cosmic scale factor. 
For standard relativistic gravity, the growth index
$\gamma=0.55$ gives an accurate description of the behaviour in
$\Lambda$CDM models. For non-standard gravity, the growth history
can be described via changes in $\gamma$ in many cases
(\citejap{LinderMG}; \citejap{PolarskiMG}). Although this is
not universally true, a useful starting point is to assume
the above relation and ask if the estimated value of $\gamma$ is
consistent with the standard 0.55. We will take this approach here.

The issues in CMB lensing tomography are similar to those in studies of
gravitational lensing using galaxy shear. The lensing distortion
of background galaxies of known redshift measures the total lensing
effect of all matter at all redshifts up to that of the background. But
if we also have foreground galaxies of known redshift, then a
cross-correlation analysis can be performed, as with the CMB
(this is known as `galaxy-galaxy lensing'). Such work has
been carried out with considerable success
(see e.g. the recent papers from KiDS and DES:
\citejap{KIDS_3x2pt2017}; \citejap{DES_3x2pt2017}).
However, tomographic lensing of the CMB has some distinct
advantages over the use of galaxy shear: the lensing estimation is
clean compared to the estimation of correlated galaxy ellipticities;
the redshift of the background CMB is known, whereas the photometric
redshifts of background lensed galaxies can introduce significant 
uncertainty. There is thus a great interest in cross-correlating 
CMB lensing with foreground galaxy structures, and some encouraging
results were obtained by the {\it Planck\/} team \citep{Planck_lens2013}
as well as from the Dark Energy Survey Science Verification data
(\citejap{Giannantonio2016}).
More recently the SDSS galaxy distribution was 
considered by \cite{Doux2017}, although they did not
use their results to constrain theories of gravity. 
The {\it Planck\/}-derived CMB lensing map was also shown to correlate with 
SDSS-based galaxy lensing \citep{Singh2017}.
Correlations have also been found between CMB lensing and
high-$z$ H-ATLAS galaxies \citep{HATLASxPlanck} and with QSOs
(\citejap{Sherwin2012}; \citejap{WISEQxPlanck}).
The precision of these results is
however limited because the catalogues under study generally
cover a smaller sky area than the all-sky {\it Planck\/} coverage, the
only exception being the WISE quasar sample.

The main aim of the present paper is therefore to carry out CMB lensing
tomography, cross-correlating the reconstructed CMB lensing map from
\cite{Planck_lens2015} with 
the largest available all-sky galaxy datasets 
with photometric redshifts (\phzs): one million galaxies from the 2MASS Photometric 
Redshift catalogue \citep{Bilicki14} and 20 million galaxies
generated by pairing the WISE survey with the SuperCOSMOS galaxy catalogue
(\citejap{Bilicki16}). We will refer to these samples as respectively 2MPZ and \WIxSC. 
Over about a quarter of the sky, these two datasets are complemented by the deeper \phz\ 
catalogue from the Sloan Digital Sky Survey (SDSS: \citejap{Beck16}); this allows for some useful cross-checks
and extension to higher redshifts.

The two all-sky datasets have already been used in
a number of analyses related to
our work. \cite{2MxPlanck} cross-correlated 2MPZ with
{\it Planck\/} lensing, and used it together with 2MPZ auto-correlations
(analysed in detail by \citealt{BABBP18}) to constrain the growth of
structure at $z\sim0.1$. \cite{Raghunathan2017}, on the other hand,
stacked {\it Planck\/} lensing convergence at positions of \WIxSC\ galaxies to
measure masses of the latter. Both these studies found significant
correlation between the samples used and CMB lensing, despite
relatively low redshifts probed by the catalogues. Our work extends
and complements these efforts, and in particular we use for
the first time \WIxSC\ and SDSS photometric samples for a tomographic
analysis of CMB lensing. The power of these datasets for
cross-correlation tomography has been already demonstrated by \cite{Cuoco17}
and \cite{Stolzner18}, where respectively Fermi-LAT extragalactic
$\gamma$-ray background and {\it Planck\/} CMB temperature fluctuations were
used as matter tracers.

We describe the 2MPZ and \WIxSC\ datasets in Section 2, together with the partner
SDSS catalogue that is used for testing of systematics and extension to
higher redshifts, with particular emphasis on the 
calibration of redshift distributions. 
The necessary elements of theory are presented in Section 3, and
the measured cross-correlations are presented in Section 4,
together with a discussion of possible systematics. 
The statistical interpretation
in terms of the growth history of density fluctuations is given in Section 5,
and Section 6 sums up and considers future prospects for such analyses.

\section{Data}

In this study, we make use of three extensive catalogues of galaxies
with \phzs. One comes from the SDSS, specifically the
DR12 photo-$z$ catalogue (\citejap{Beck16}; see also \citejap{Reid2016}).
The two others are shallower, but cover nearly three times the sky area of 
the SDSS. These involve combining longer-wavelength data with legacy
optical photographic photometry from the SuperCOSMOS all-sky galaxy
catalogue \citep{Hambly01a,Hambly01b,Peacock16}. The most accurate results come from
joining this information with the near-infrared (IR)
data from the Two Micron All Sky Survey
\citep[2MASS:][]{Skrutskie06} Extended Source Catalogue
\citep[XSC:][]{Jarrett00},
together with 3.4- and 4.6-micron
photometry from the Wide-field Infrared Survey
Explorer \citep[WISE:][]{Wright10}. 
A considerably deeper \phz\ dataset is obtained by using SuperCOSMOS and WISE only.

The 2MASS Photometric Redshift
catalogue\footnote{Available for download from
  \url{http://ssa.roe.ac.uk//TWOMPZ}.} \citep[2MPZ:][]{Bilicki14} was
constructed by matching the 2MASS XSC with both SCOS and WISE.
As the two latter datasets are
much deeper than 2MASS, such a cross-match was successful for a vast
majority of the sources ($>95\%$ in unconfused areas) and provided
8-band photometry for all the matched galaxies, spanning from
photographic $B R I$, through near-IR $J H K_s$ up to
mid-IR $W1$ and $W2$. Spectroscopic redshifts available for a large
subset of all 2MASS (over 30\%) from such surveys as SDSS, 6dFGS, and
2dFGRS provided a comprehensive calibration set for deriving \phzs\ for all the 2MPZ sources, using the ANNz
artificial neural network tool \citep{ColLah04}. After applying a flux
limit of $K_s<13.9$ (Vega) to ensure uniformity of the coverage, the
final 2MPZ sample includes 940,000 galaxies on most of the sky (except
for very low Galactic latitudes). Its median redshift is $\langle z
\rangle = 0.08$ and the typical \phz\ error $\sigma_{\delta
  z}\simeq0.015$. In principle, 2MPZ should be reliable to lower
Galactic latitudes than \WIxSC, but for simplicity we made a
conservative choice and applied the \WIxSC\ mask to the 2MPZ data. 
This gives $\simeq70\%$ of sky available for this analysis.

\begin{figure}
\centering
\includegraphics[width=0.5\textwidth]{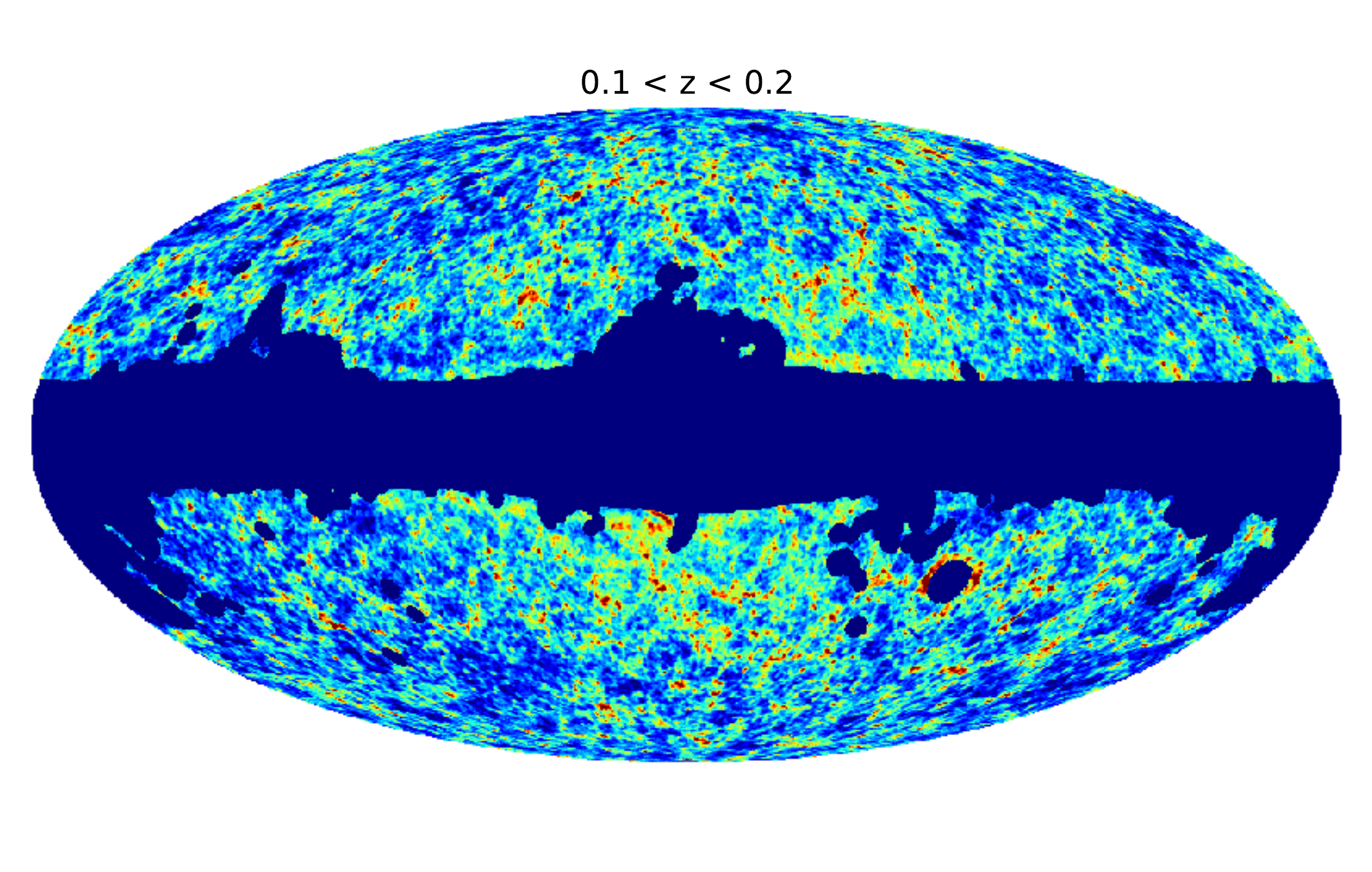}
\vglue -2em
\includegraphics[width=0.5\textwidth]{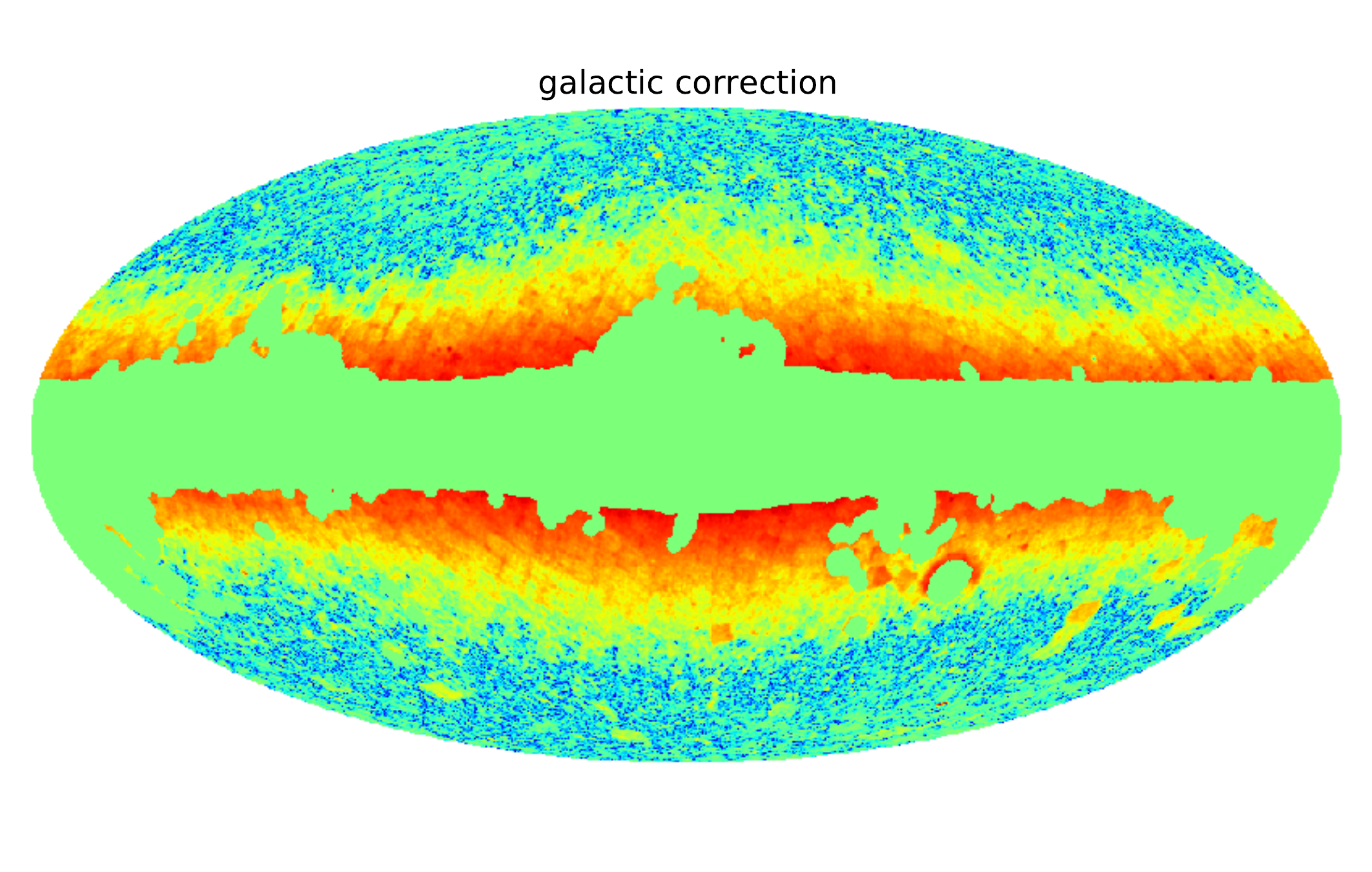}
\vglue -2em
\includegraphics[width=0.5\textwidth]{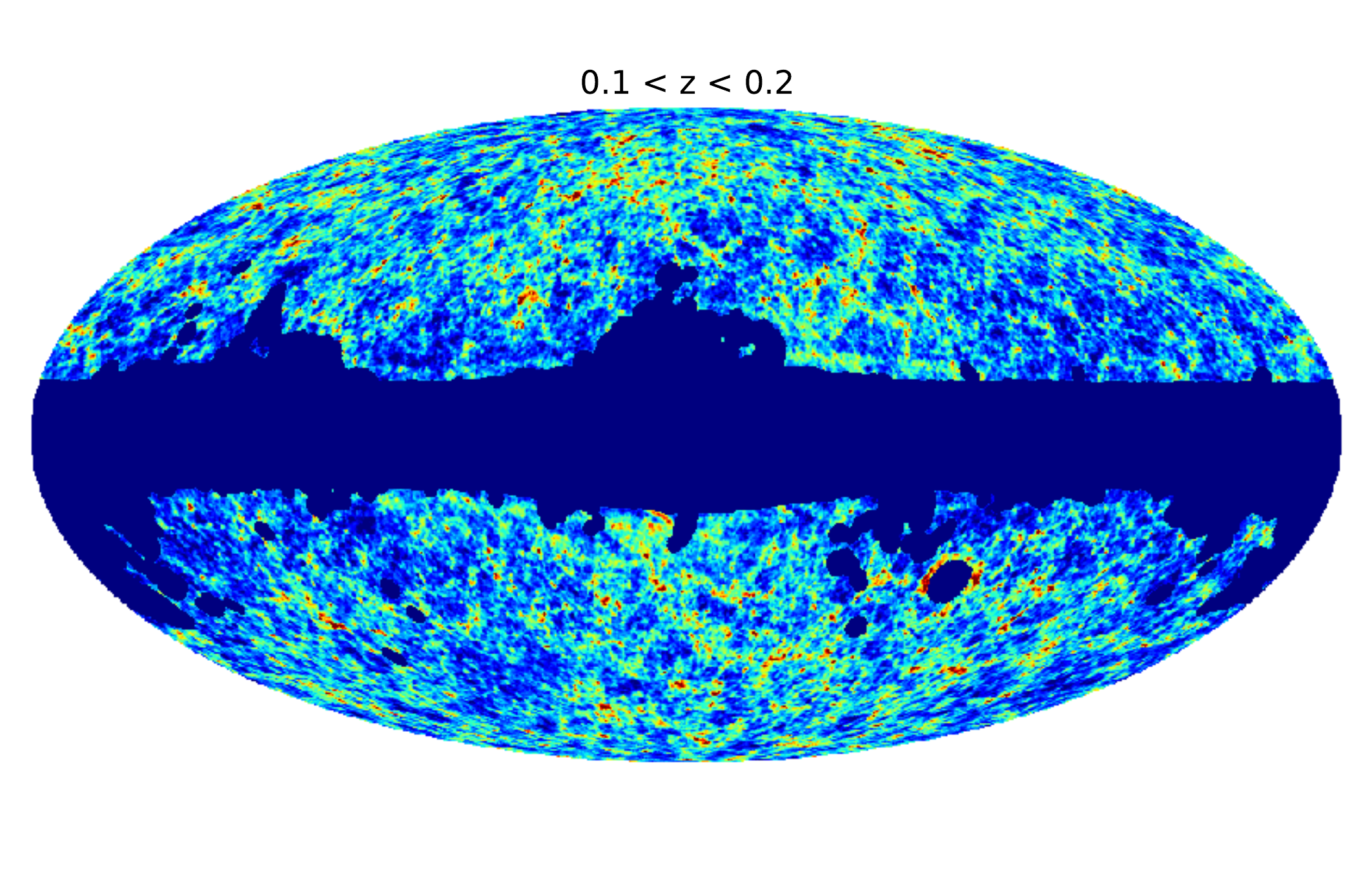}
\caption{Illustrating the removal of residual stellar contamination
from one of the \WIxSC\ tomographic slices by correlating galaxy
density with total (star-dominated) WISE surface density. Only the
$0.1<z<0.2$ slice is illustrated, but the same procedure was applied to all
slices. The top image shows the galaxy surface density smoothed with 0.5$^\circ$ FWHM,
stretched between 0.5 and 2 times the mean density. The middle
image shows the inferred stellar contamination, as described in the text,
and the final image shows the corrected data.
}
\label{fig:starclean}
\end{figure}

The WISE $\times$ SuperCOSMOS photometric
redshift catalogue\footnote{Available for download from
  \url{http://ssa.roe.ac.uk/WISExSCOS}.} \citep[\WIxSC:][]{Bilicki16}
is an extension beyond 2MPZ made by combining SCOS and WISE only. These
two datasets provide a much deeper ($\sim3\times$) galaxy sample than
possible with 2MASS, and with over 20 times larger surface density,
although its sky coverage useful for extragalactic science is smaller,
about 70\% of sky after applying the relevant masks
(see \citejap{Bilicki16} for details of the construction of the
\WIxSC\ mask). \WIxSC\ includes
almost 20 million galaxies with $\langle z \rangle = 0.2$ but having a
broad $dN/dz$ reaching up to $z\sim0.4$. The \phzs\ were derived based
on four bands $(B,R,W1,W2)$, again employing the ANNz package.
In practice, the neural
networks were trained on a complete calibration set from the
equatorial fields of the Galaxy And Mass Assembly
\citep[GAMA:][]{Liske15} survey. As in the 2MPZ case, these
\phzs\ exhibit minimal mean bias but have larger average scatter of
$\sigma_{\delta z/1+z} = 0.035$, as expected due to the availability
of half as many bands in \WIxSC\ as compared to 2MPZ.

One distinct issue affecting the \WIxSC\ dataset is stellar
contamination. Despite masking areas of high stellar density and
applying appropriate colour cuts, some stellar contamination remains
at the few per cent level -- a significantly larger issue than
for the 2MPZ or SDSS samples. This contamination results largely
from stellar blends producing spurious extended sources, and is
thus concentrated towards the Galactic plane. Moreover, because of the
distinct colours of such objects, the effects tend to be concentrated
in particular \phz\ slices. We deduced a correction for this effect
by plotting `galaxy' surface density against the total WISE
surface density (a good proxy for the stellar surface density),
fitting a smooth nonlinear relation, and subtracting 
a version of the WISE total surface density map with the counts modified
using this nonlinear relation, in order to predict the angular variation of
the stellar contamination.
The result of this process is shown in Fig. \ref{fig:starclean},
and can be seen to yield cosmetically cleaner tomographic slices.
In practice this cleaning has negligible quantitative impact on the
cross-correlation analysis, since it affects only the lowest
angular wavenumbers; nevertheless, this was an important check to carry out.

The DR12
\phz\ catalogue\footnote{\url{http://www.sdss.org/dr12/algorithms/photo-z/}}
\citep{Beck16} is the most recent such dataset available from the SDSS,
superseding earlier samples of that kind. Photo-$z$s and their error
estimates, together with specific quality classes, were derived for
about 200 million galaxies from the SDSS photometric catalogue, using
the local linear regression technique based on a spectroscopic
calibration set composed mostly of galaxies from the SDSS DR12
spectroscopy, plus several other surveys.
The estimated precision of these \phzs\ varies, and \cite{Beck16}
recommend filtering according to error classes and using only
sources of class 1 and perhaps also $-1$, 2 and 3. After careful
inspection of the catalogue, we have however decided not to follow
these recommendations in order to guarantee uniform selection 
of the sample over the sky,
and to maximise its surface density. The assignment to the particular
classes is based on errors in the original SDSS photometry; hence, the
variations in these classes are strongly correlated with the quality
of observations, changing from pointing to pointing. Indeed, we have
verified that if we followed the recommendations to preserve only
those particular classes, the resulting sample would exhibit
significant variations in depth and surface density following the SDSS
scanning pattern. Moreover, a sample preselected according to the
recommended classes would include only 55 million sources out of the
200 million available in total. We have thus decided to accept
a poorer redshift quality in exchange for improved sampling uniformity and
density. The only cut we apply on the parent sample (in addition to
masking) is to require an estimate of the photo-$z$ to be
given; this preserves over 185 million galaxies with
$\langle z \rangle =0.44$ and $dN/dz$ reaching up to $z\sim1$. In our
analysis we however use only sources in the range of $0.1<
\zph <0.6$: the more distant shells have a lower number density,
making the cross-correlation results noisy, and it is harder
to check the calibration of the distribution of true
redshifts for photo-$z$ data in this regime (see section \ref{sec:dndz}).
The SDSS photometry in DR12 were obtained for a
number of distinct sub-projects. In order to focus on a consistent extragalactic
dataset, we applied a mask corresponding to the 9367 deg$^2$ of the 
BOSS project \citep{Reid2016}.

\begin{figure}
\centering
\includegraphics[width=0.48\textwidth]{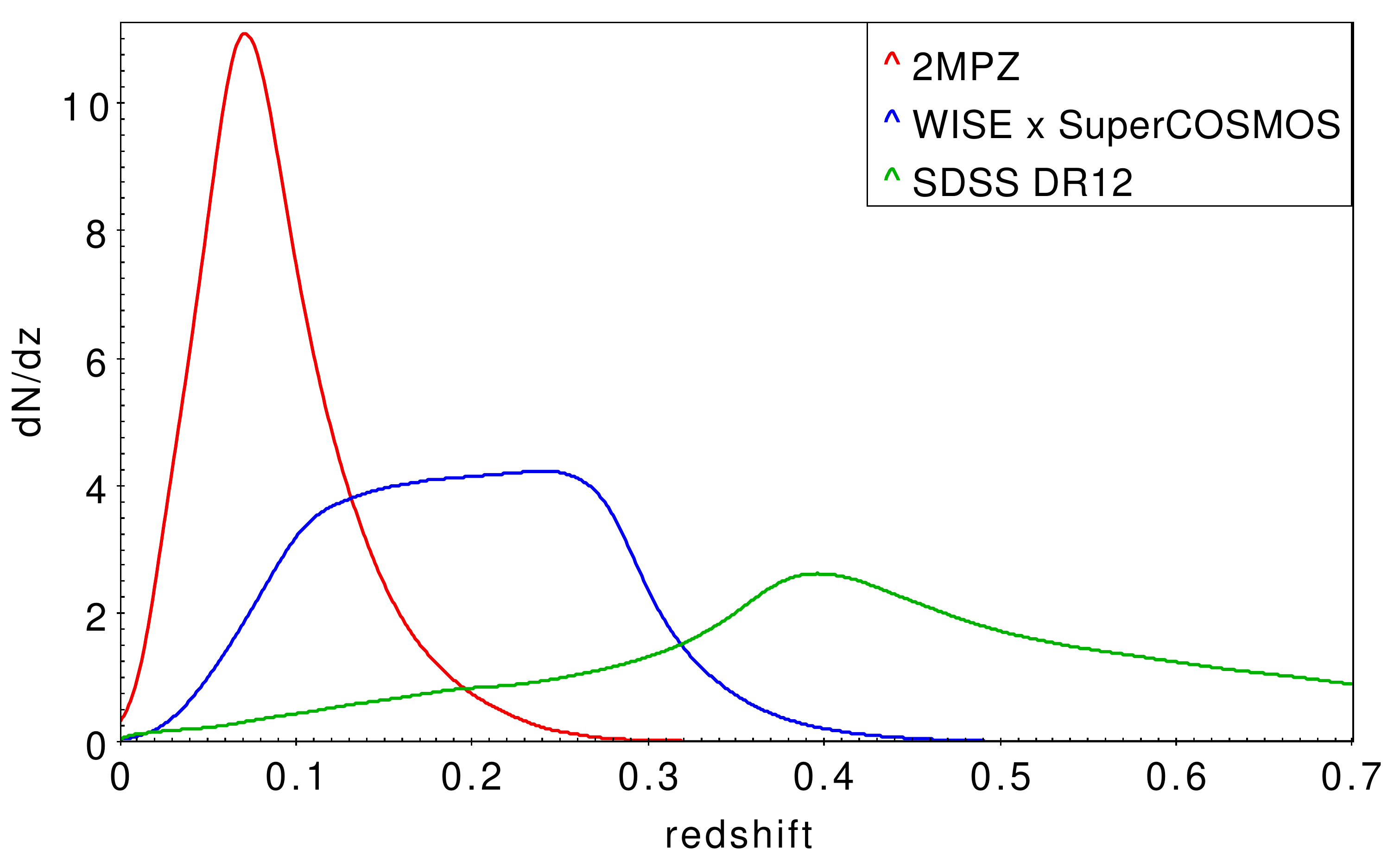} 
\caption{Normalized photometric redshift distributions of the three catalogues used in this study.
}
\label{fig:dNdz_catalogues}
\end{figure}

\begin{figure*}
\centering
\includegraphics[width=1.0\textwidth]{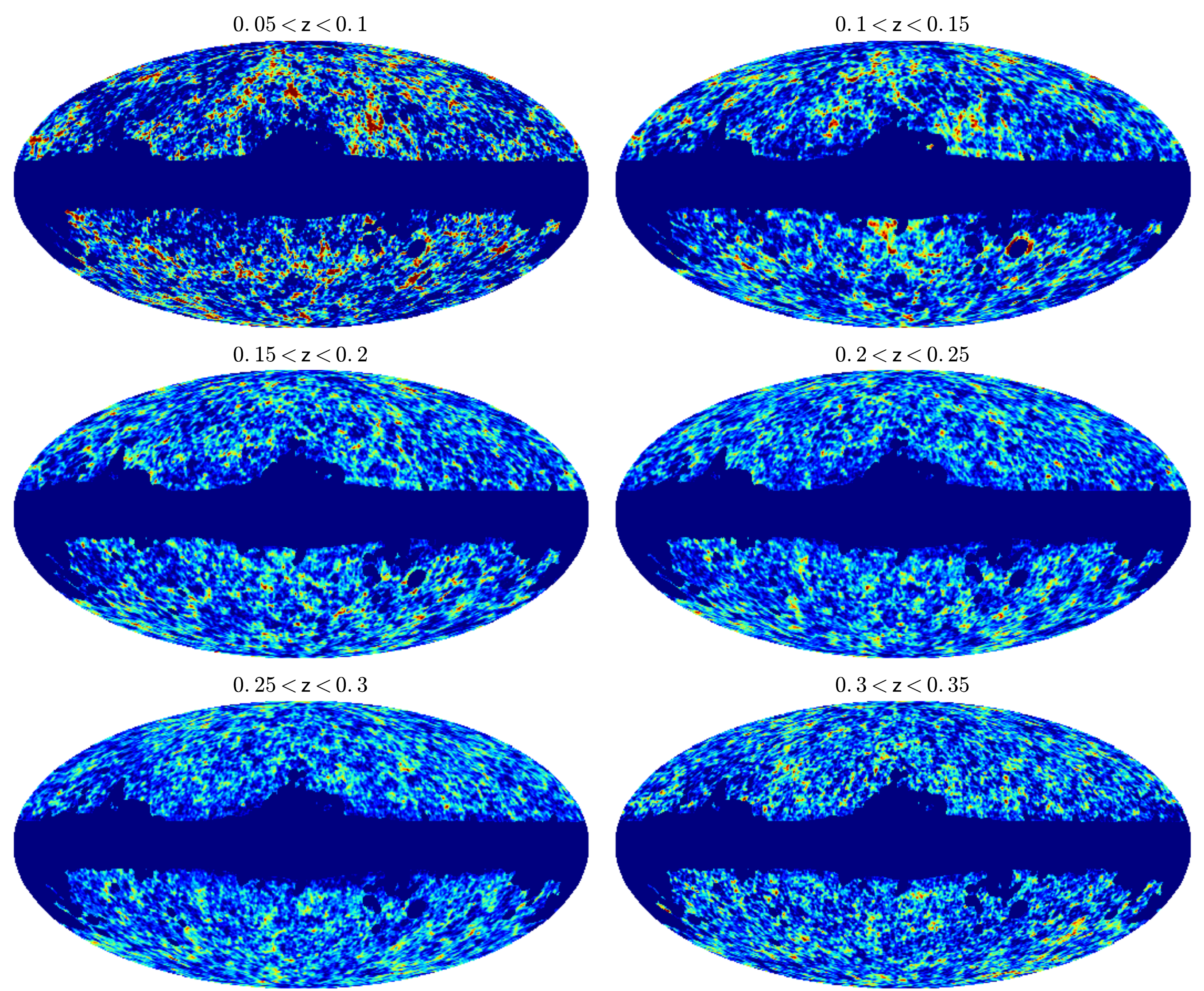}
\caption{The galaxy surface density in various \phz\ slices for 
the 2MPZ ($0.05<z<0.1$) and \WIxSC\ (other $z$ bins),
The number counts have been
smoothed with a 1 degree FWHM Gaussian, and the colour scale
spans 0.7 to 2 times the mean density. The \WIxSC\ slices have
been corrected for residual stellar contamination as illustrated in Fig. \ref{fig:starclean}.}
\label{fig:sky_wisc}
\end{figure*}

\begin{figure*}
\centering
\includegraphics[width=1.0\textwidth]{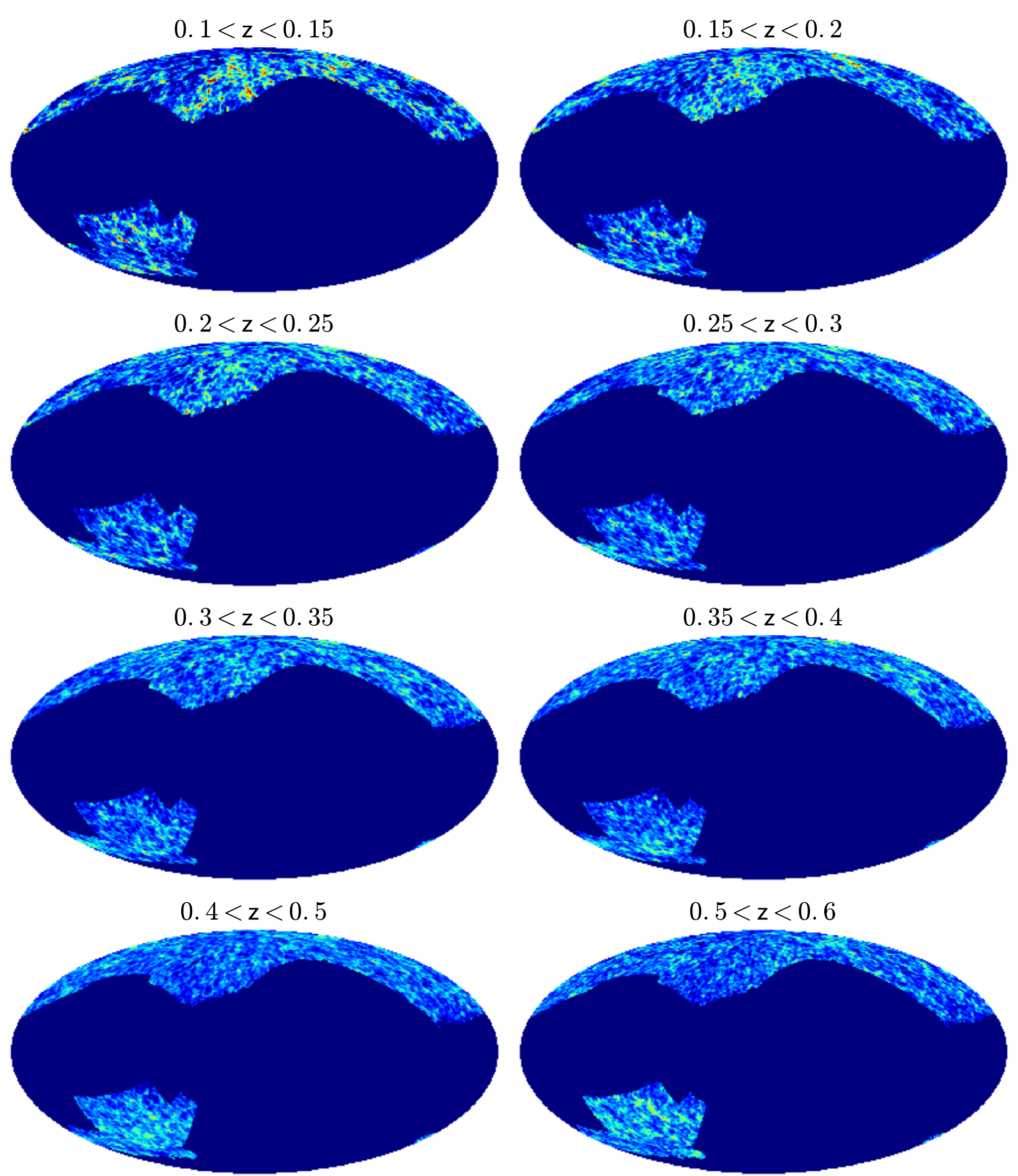}
\caption{The equivalent of the previous figure, now using SDSS \phz\
data (for which no stellar correction was made). Again,
the number counts have been
smoothed with a 1 degree FWHM Gaussian, and the colour scale
spans 0.7 to 2 times the mean density. The redshift range
is larger than for \WIxSC, but the sky coverage is less.
Although the comparison is not exact, because the SDSS \phz\
precision is higher, a visual inspection reveals similar LSS features
in both sets of slices. This agreement is quantified in the text.
}
\label{fig:sky_sdss}
\end{figure*}

In Fig. \ref{fig:dNdz_catalogues} we show the \phz\ distributions of
the three samples used in this study. Dividing this into tomographic
slices is relatively arbitrary: too coarse binning limits the chance
to study any evolution with redshift, but too narrow shells will
lack a clear signal. In practice, we were guided by the precision
of the photo-$z$s, discussed in the next section, and opted for 
6 bins between $z=0.05$ and $z=0.35$ for the all-sky (2MPZ+\WIxSC) data, and 8
bins between $z=0.1$ and $z=0.6$ for the SDSS data. We ignore
extremely local volumes where the fractional \phz\ errors become large.
The sky distribution in these 14 tomographic slices is shown
in Figs \ref{fig:sky_wisc} \& \ref{fig:sky_sdss}.

\subsection{Redshift distributions}
\label{sec:dndz}

For accurate theoretical predictions, we require a good model for
the probability distribution of true spectroscopic redshifts that
results from a given \phz\ selection, $p(z)$.
Photometric redshifts are
generally calibrated so that the distribution of true $z$ at given
$z_{\rm phot}$ is unbiased -- but we still need to know the scatter in
order to provide the appropriate broadening of $p(z)$.

We derive the redshift distributions of our photometric samples
using overlapping spectroscopic datasets. For 2MPZ and \WIxSC, this is
done with the same data which were employed for the \phz\ training. In
the former catalogue this is mostly the SDSS Main Sample, while in the
latter it is stage II of the Galaxy And Mass
Assembly (GAMA) survey \citep{Liske15}.
For the SDSS photometric sample we also employed
GAMA for \phz\ calibration, it is however too shallow to probe the
full depth of SDSS which we use here ($z<0.6$); we thus added also
information from deeper SDSS spectroscopic samples 
available in Data Release 13, 
although we note that beyond the main
sample of $r<17.77$, the SDSS spectroscopic galaxies are sparsely
sampled with specific colour preselections, which makes this dataset
very incomplete and generally biased as a \phz\ calibrator.

GAMA-II covers several fields, of
which three equatorial ones have a very high ($98.5\%$) spectroscopic
completeness down to $r<19.8$ and include almost 200,000 galaxies at
$z<0.6$ with a median $\langle z \rangle \simeq0.2$.  SDSS DR13 spectroscopic
\citep{SDSS_DR13} includes over 2.6 million galaxies at redshifts
$z<1$, albeit not with the simple magnitude-limited selection seen in GAMA.

In order to allow for flexibility, to derive estimates of $dN/d\zsp$
for each \phz\ bin we modelled the conditional distributions of the
true redshift at given photometric redshift, $\delta z \equiv \zsp - \zph$, 
as a function of
redshift. Then the true redshift distributions for each bin are
estimated from the photometric ones via
\[
p_s(z_s) = \int p_p(z_p)\, p_{\delta z}(z_s - z_p)\, \de z_p ,
\]
where subscript `$s$' stands for spectroscopic and `$p$' for
photometric.  
Note that this is not the same as considering the
`photo-$z$ error distribution', which would be the
conditional distribution of photo-$z$ at given true spectroscopic
redshift. This is in general biased, so that
$\langle (z_p|z_s) \rangle \ne z_s$, whereas the mean true
redshift at given $z_p$ should be unbiased by construction.
Certainly, what we require here is the conditional
distribution of $z_s$ at given $z_p$: this allows us
to construct the distribution of true redshifts that
arises when we make a particular photometric selection.

The redshift difference distributions $p_\delta$ were calibrated
with the aforementioned spectroscopic data, which yields
an empirical error distribution based on all the
objects of known redshift in any photo-$z$ bin. But to avoid
binning, it is convenient to use a model for the error distribution,
and we considered two options: Gaussian and modified Lorentzian. The
latter takes the form
\[
\label{eq:Lorentz}
p_\delta \propto \left(1+\frac{\delta z^2}{2 \, a \, s^2} \right)^{-a}\;,
\]
where $a$ and $s$ are fitted parameters;
such a formula represented the 2MPZ redshift residual distribution
very well (see \citealt{Bilicki14}).
 This flexible form allows us to
account for both non-Gaussian wings as well as a more peaked
distribution at the centre, both often being characteristics of
\phz\ error distributions.

In the modelling we tested various levels of sophistication.  As
mentioned above, the \phzs\ we use are constructed so that they are to
a good approximation unbiased in the mean value of true $z$ at a given
\phz; thus we always assume the mean of $\delta z$ to be 0 at any
redshift, and all the interest lies in the distribution of $\delta z$.
The simplest yet least realistic model is to assume a Gaussian $\delta
z$ with scatter evolving linearly with redshift, i.e.
redshift-independent $\delta z/(1+\zsp)$. This can be then extended to
a more general form of $\sigma(z)$ but still assuming a Gaussian form. 
A more accurate description is obtained
by using the generalised Lorentzian
\eqref{eq:Lorentz}, which has two free parameters controlling the
scatter and the wings; here assuming linearly evolving $a(z)$ and
$s(z)$ is general enough to capture the actual $\delta z$
behaviour. Finally, in the case of the SDSS \phzs, \cite{Beck16}
provide estimates of rms $\delta z$ for each source, so we were able to
test a model where these individual errors were used to estimate the true
redshift distributions from the photometric ones. In general,
this approach did not give a good agreement with the direct
inference of the $\delta z$ distribution using calibrating
spectroscopy; this probably reflects the fact that the SDSS
error estimates are only claimed to be reliable
for class 1 sources (see \citealt{Beck16} for details), whereas
we make no selection on class.

For \WIxSC\ the best-fit Gaussian scatter is $\sigma(z) = 0.08 z +
0.02$ while the Lorentzian parameters were fitted individually for
each redshift bin; their redshift dependence is approximately $a(z)
\simeq -4 z + 3$ and $s(z) \simeq 0.04 z + 0.02$. In the case of SDSS,
calibration on SDSS spectroscopy only (dominated by LRGs) gives
best-fit Gaussians with $\sigma(z) = 0.02(1+z)$; more realistic
modelling with GAMA (+SDSS at $z>0.4$) indicates $\sigma(z) =
0.03(1+z)$; finally, if the published \phz\ errors from \cite{Beck16}
are used, then the overall $p_\delta$ can be approximated with a
Gaussian of $\sigma(z) = 0.05(1+z)$. Of these three models we consider
the middle one to be the most realistic. 
For the SDSS data, we did not find that it was necessary to
resort to non-Gaussian error distribution models in order
to obtain a good description of the results; as might have been
expected, the digital SDSS photometry is better defined and
less subject to outliers than the SCOS legacy photographic measurements.

\begin{figure}
\centering
\includegraphics[width=0.5\textwidth]{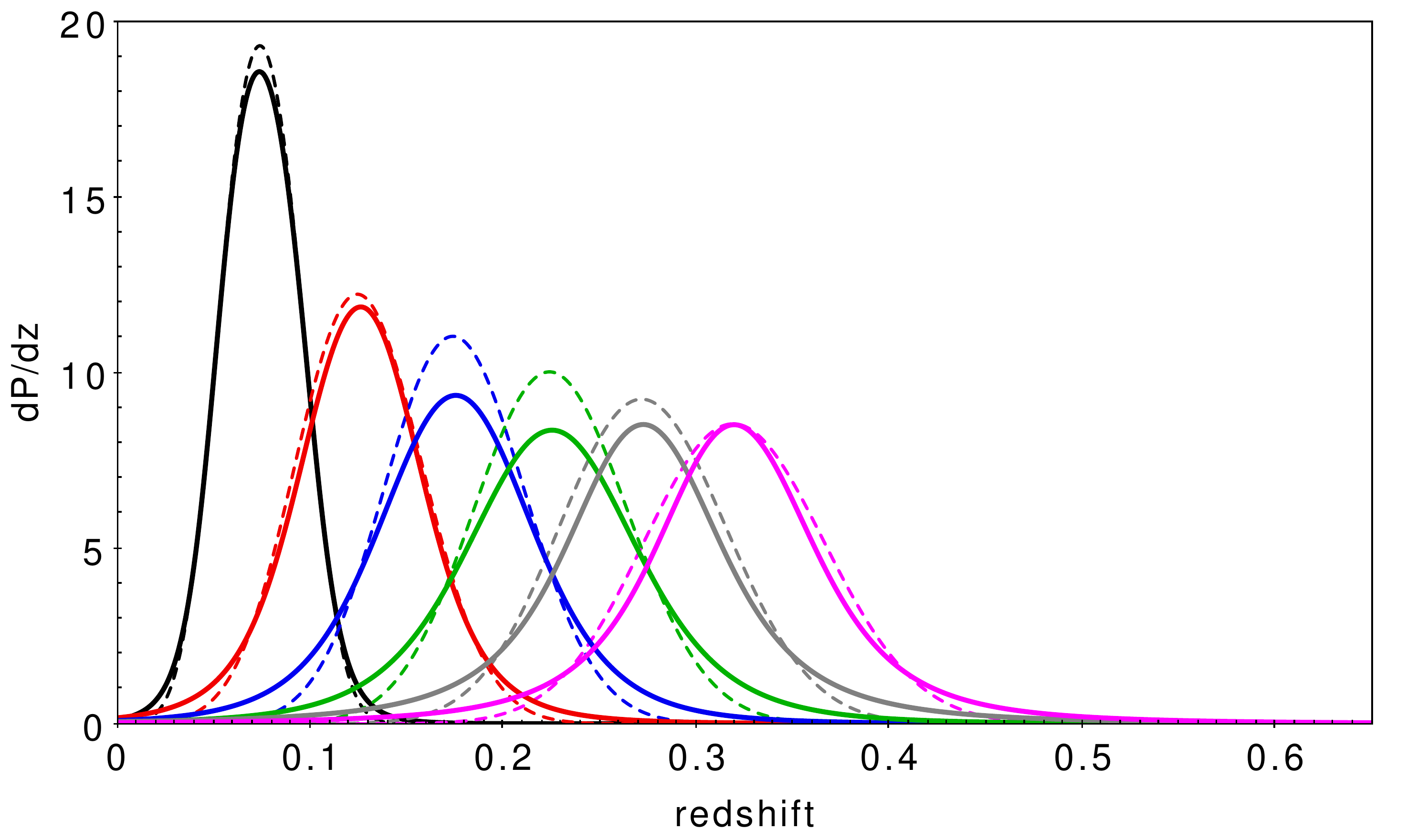} 
\includegraphics[width=0.5\textwidth]{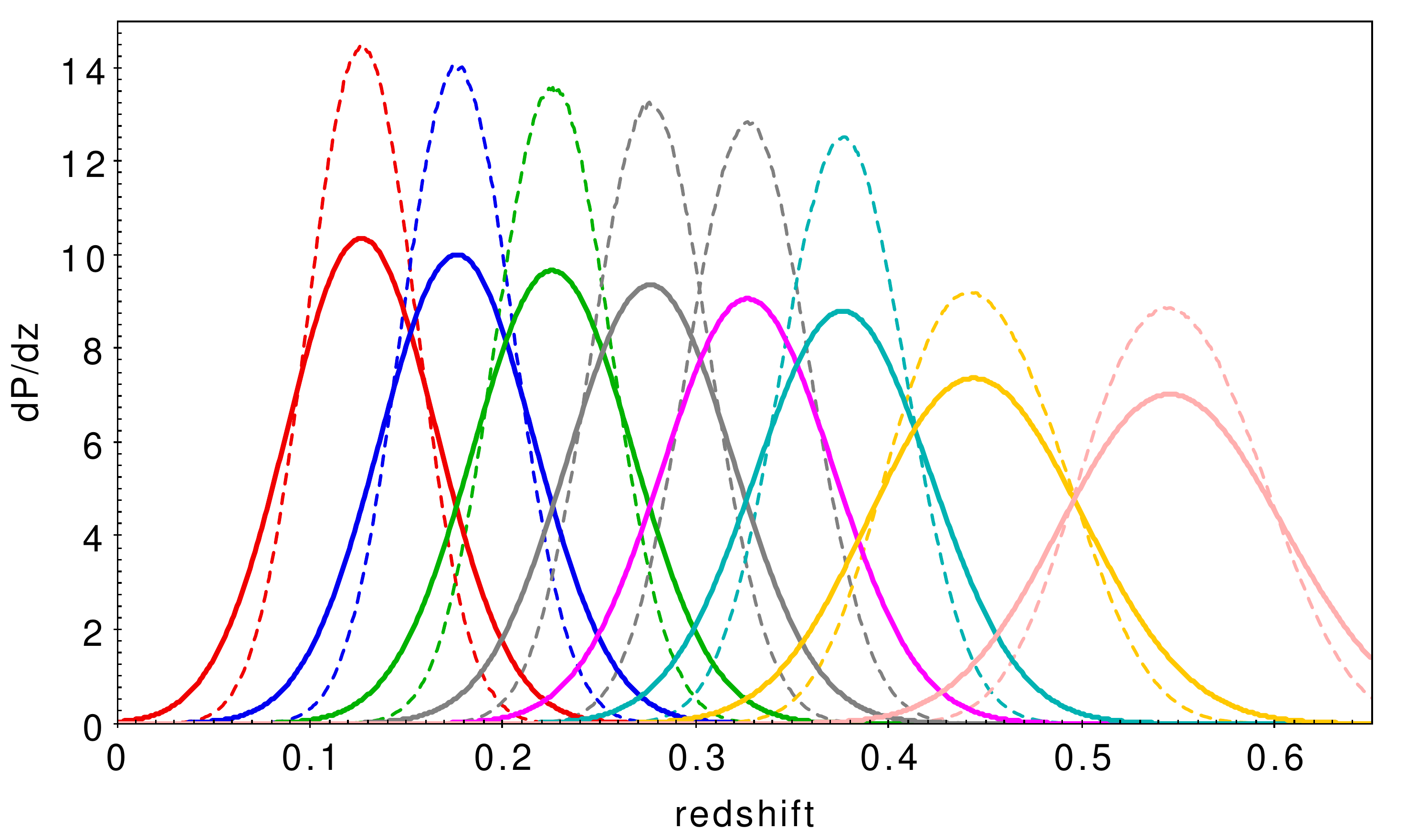} 
\caption{Calibrated redshift distributions of the 2MPZ+\WIxSC\ (upper) and
SDSS  DR12 catalogues (lower), derived by convolving the \phz\ distribution
in each slice with an error distribution.
In the top panel, dashed curves use a Gaussian convolving function and solid
use a modified Lorentzian (adopted model).
Black lines show the distributions for 2MPZ (single bin).
In the lower panel, the lines are
for two different Gaussians: dashed is $\sigma=0.02$ and
solid is $\sigma=0.03$ (adopted model).}
\label{fig:dNdz_bins}
\end{figure}

As far as 2MPZ is concerned, we used only one redshift bin, and we
modelled its $p_\delta$ as either a Gaussian or a modified
Lorentzian with redshift-independent parameters, in a similar way as
was done for the whole 2MPZ sample in \cite{Bilicki14}. As the bin we
use for 2MPZ, $0.05<\zph<0.1$, includes over half of all the 2MPZ
galaxies and is centred almost on the median redshift of the sample,
it is not surprising that the best-fit parameters of the model are
here very similar to those obtained in \cite{Bilicki14}; namely, for
the Gaussian $\sigma_{\delta z} = 0.014$ and for the Lorentzian,
$a=2.93$ and $s=0.012$. See also \cite{BABBP18} for a recent validation 
of 2MPZ \phz\ performance and of the catalogue itself.

The resulting estimates of $dN/d\zsp$ are shown in Fig.~\ref{fig:dNdz_bins}.
The impact of the different model choices on the lensing predictions
are discussed below in section \ref{sec:robustness}.

\begin{figure}
\centering
\includegraphics[width=0.48\textwidth]{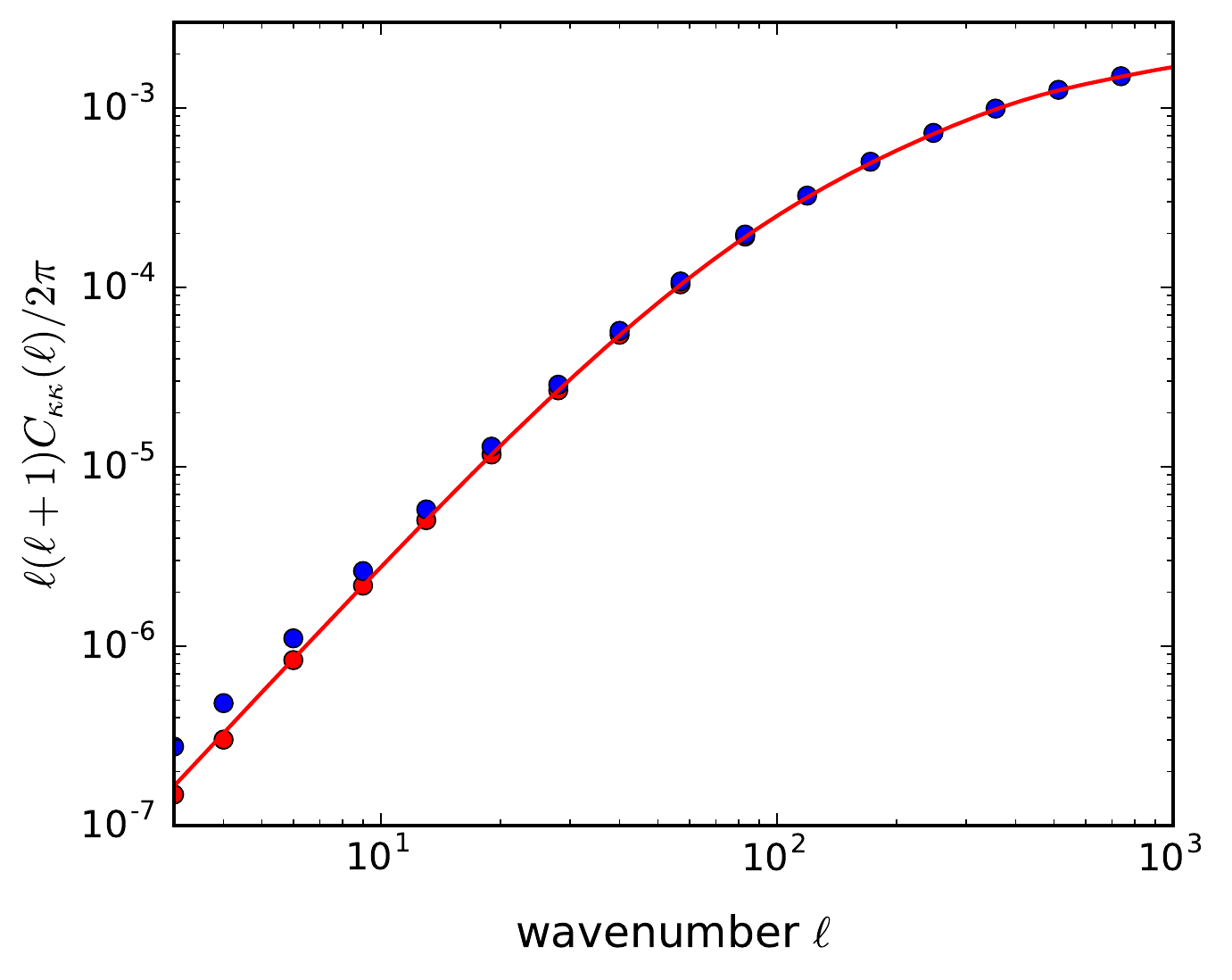}
\includegraphics[width=0.48\textwidth]{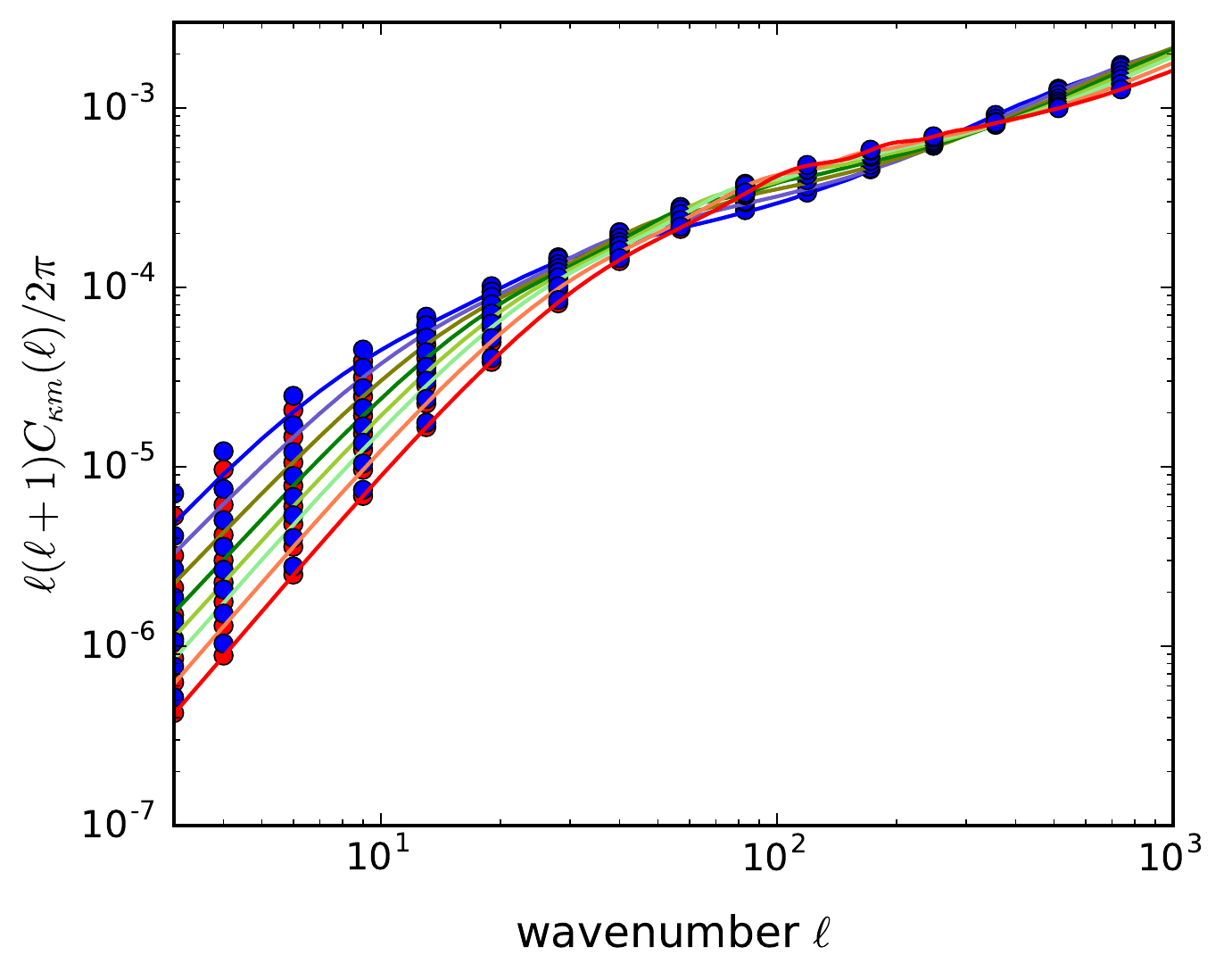}
\includegraphics[width=0.48\textwidth]{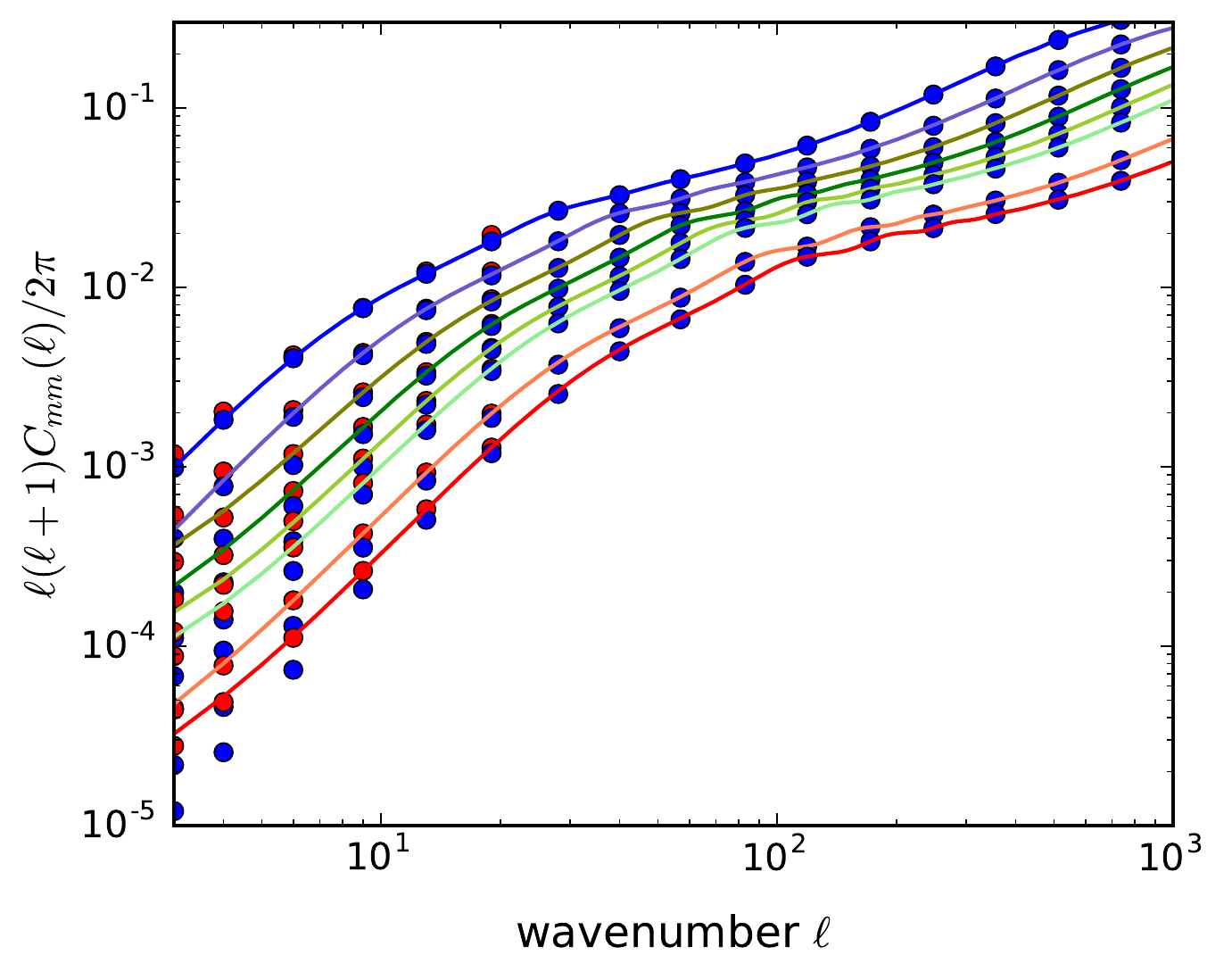}
\caption{Theoretical predictions for the various power spectra.
These plots are shown for the case
of eight SDSS bins in photometric redshift,
ranging between mean redshifts of 0.125 to 0.55. 
Red points are exact integration; blue points are the Limber--Kaiser
approximation; lines show the interpolated adopted results,
colour-coded from
blue (lowest $z$) to red (highest $z$); 
the signals tend to
decrease with increasing redshift.
The panels show, in order: lensing auto-power;
lensing-mass cross-power; mass auto-power.
}
\label{fig:pows_theory}
\end{figure}

\begin{figure}
\centering
\includegraphics[width=0.48\textwidth]{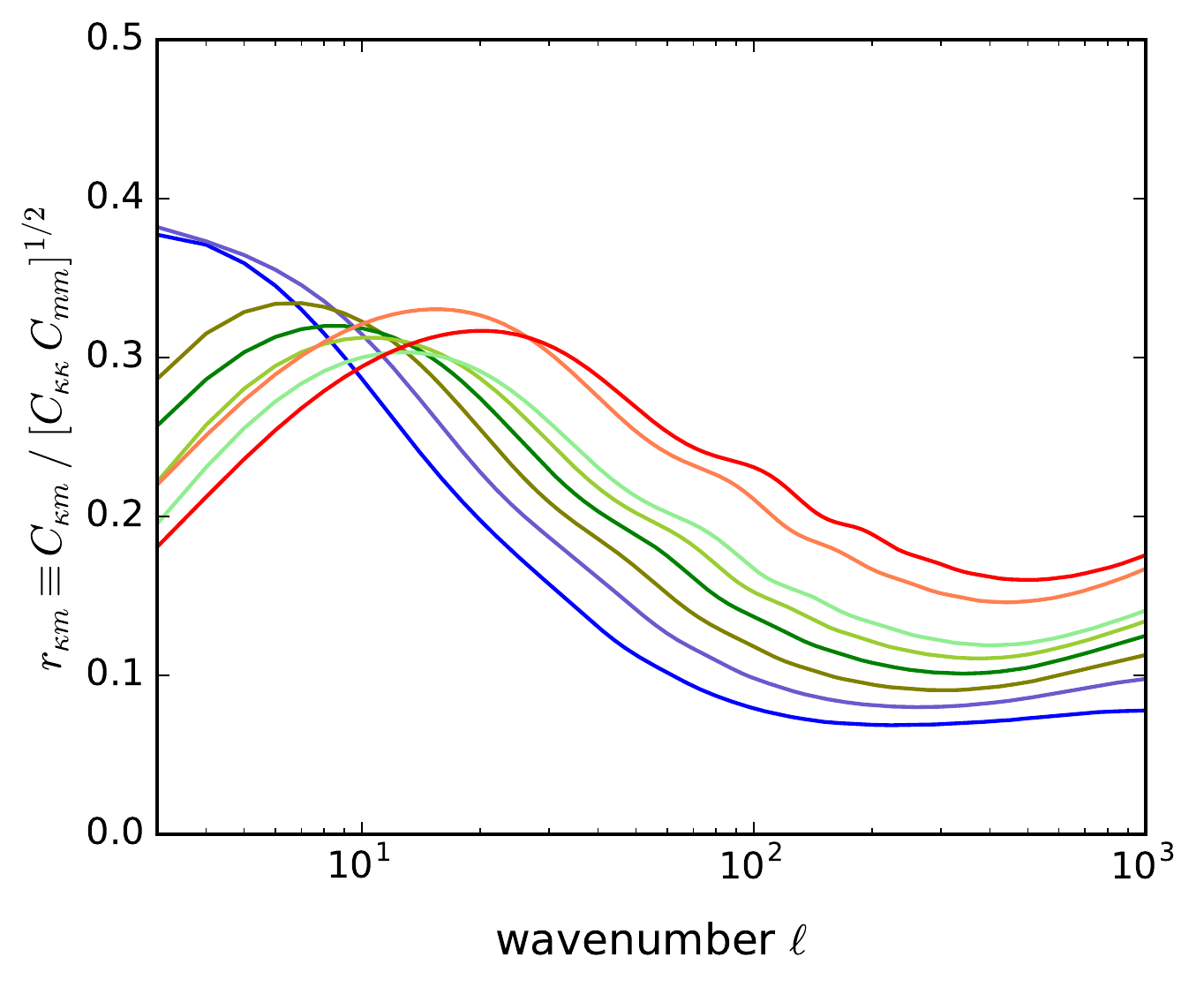}
\caption{The harmonic-space correlation coefficient, $r_{\kappa m}$, between lensing and 
clustering \eqref{eq:rmk}, derived from the predictions of Sec. \ref{sec:theory}.
This is shown for the case of the eight SDSS bins in photometric redshift,
ranging between mean redshifts of 0.125 to 0.55. The lines are colour-coded from
blue (lowest $z$) to red (highest $z$).
Increasing redshift moves the peak to higher $\ell$ and raises
the amplitude at high $\ell$. 
}
\label{fig:corr_theory}
\end{figure}

\section{Theory}
\label{sec:theory}

The theory of CMB lensing is reviewed comprehensively by
\cite{LewisChallinor2006}, and we summarise the key elements here.
The gravitational-lens deflection is the 2D angular gradient of a potential,
$\boldmath{\alpha} = \boldmath{\nabla}\psi$, where the lensing potential $\psi$
is related to the convergence, $\kappa$, via $\nabla^2\psi = 2\kappa$
(note that \citealt{LewisChallinor2006} define $\psi$ with the opposite sign).
Thus, in terms of angular power spectra,
\[
C_{\kappa\kappa}=[\ell(\ell+1)]^2 C_{\psi\psi}/4.
\]
For a flat universe (assumed here), the CMB convergence is a projection of the 
fractional density fluctuation, $\delta$, with a density-dependent kernel:
\[
\kappa = {3 H_0^2\Omega_m\over 2 c^2}
\int_0^{r_{\rm LS}}dr\;  \delta(r)\; {r(r_{\rm LS}-r)\over a\, r_{\rm LS}}
\equiv \int \delta(r)\; K(r)\; dr.
\]
Here, $r$ is comoving distance to the element of lensing matter, and this is
integrated from the origin to the last-scattering surface at $r_{\rm LS}$.

For two quantities, $a$ and $b$,
obeying similar relations with kernels $K_1$ and $K_2$,
the angular cross-power at multipole $\ell$ is
\[
C_{ab}(\ell) = 4\pi
\int\Delta^2(k)\; d\ln k\; \int K_1 j_\ell(kr)\, dr \; \int K_2 j_\ell(kr)\, dr,
\label{eq:pproj}
\]
where $k$ is the comoving spatial wavenumber, $\Delta^2(k)$ is the
dimensionless matter power spectrum, and $j_\ell$ is a spherical
Bessel function.  In the large-$\ell$ limit, the Bessel functions
become sharply peaked, and we obtain a cross-power version of Kaiser's
harmonic-space version \citep{Kaiser1992} of the \cite{Limber} equation:
\[
C_{ab}(\ell) ={\pi\over\ell}\int \Delta^2(\ell/r)\, r\, K_1(r)K_2(r)\, dr.
\]
For galaxy data, the kernel must represent the probability distribution of redshift,
$p(z)$, thus $K(z) = p(z) (dr/dz)^{-1}$. 

If we initially neglect galaxy bias, and assume that mass can be probed directly
with the same $K(z)$ kernel, then the 
relevant power statistics are those relating the total CMB lensing
optical depth ($\kappa$) and the projected mass ($m$) overdensity in a given
\phz\ slice ($m$): $C_{\kappa\kappa}(\ell)$, $C_{\kappa m}(\ell)$
and $C_{mm}(\ell)$. These are illustrated
in Fig. \ref{fig:pows_theory} for the SDSS \phz\ bins adopted here;
the low-$z$ \WIxSC\ curves are very similar in form.
These plots show that the Kaiser-Limber approximation is indistinguishable
from the exact projected power of eq. (\ref{eq:pproj}) for $\ell\gs 20$; we therefore
use this approximation at the higher multipoles where it is numerically faster
and more stable than the exact expression.

It is interesting to combine these auto- and cross-power measurements
into a harmonic-space correlation coefficient
\[
r_{\kappa m} \equiv C_{\kappa m}\,/\,(C_{mm}C_{\kappa\kappa})^{1/2},
\label{eq:rmk}
\]
which has the interpretation that $r_{\kappa m}^2$ gives the fraction
of the total CMB lensing variance that is contributed by the tomographic
slice under consideration. This quantity is plotted in Fig.
\ref{fig:corr_theory}, where it can be seen that typical figures
are 0.3 at $\ell\simeq 10$, declining to between 0.1 and 0.25 at
$\ell=100$, for bins of width $\Delta z =0.05$ out to $z\sim1$.
Thus such tomographic bins can capture a significant fraction of the total lensing variance
at $\ell=10$, but only a few per cent of the total variance at $\ell=100$.

\subsection{Fiducial model and nonlinearities}

These theoretical predictions require a choice of cosmological parameters. The best
choice of these remains subject to slight debate concerning `tensions' between {\it Planck\/}
and other determinations. Our reading of the situation is that there is no
conclusive evidence that any of the main determinations are in error by
more than their reported statistical uncertainties. Motivated by
\cite{Planck_pars2015}, \cite{KIDS_3x2pt2017} and \cite{DES_3x2pt2017},
we adopt the following flat $\Lambda$CDM model:
\[
(\Omega_m, \Omega_b, h, \sigma_8, n_s) = (0.3, 0.045, 0.7, 0.8, 0.965). 
\]
The remaining uncertainties in these parameters are all at the level of 2--3\% 
(or 0.5\% in $n_s$), which constitutes a negligible variation in the context of
the current precision of the data presented here. Thus we will 
generally treat the $\Lambda$CDM
predictions as specified perfectly by this simple fiducial model.
We will then want to see how well the fiducial cross-correlation predictions
agree with the measured signal. Any mismatch in amplitude can be interpreted
as requiring a change in the evolving amplitude of mass fluctuation, $\sigma_8(z)$.
A key aspect of the current analysis is that the fiducial 
$\sigma_8(z)$ is an extrapolation assuming the growth rate from
standard gravity; the measured growth history
inferred from the tomographic data can therefore be used to set limits on deviations
from this rate. In practice, we will use the growth-index
parametrisation, $f_g=\Omega_m(a)^\gamma$ to capture this information.

Although the lensing is weak, this does not mean that only linear scales
are probed. In practice, we will work to angular multipoles of $\ell=300$,
at redshifts of typically 0.2, corresponding to wavenumbers $k=0.5\hompc$;
nonlinear corrections are significant at these scales. We estimate these
corrections using the HALOFIT code of \cite{halofit}. More recent work
has shown that the CDM simulations used to calibrate the method were
systematically low in small-scale power, and revised fits were produced
by \cite{halofit2}. A simple alternative of comparable accuracy is to
correct the original predictions by the following factor,
in a manner that is taken to be independent of redshift:
\[
(P-P_{\rm lin}) \rightarrow (P-P_{\rm lin}) \times (1+2y^2)/(1+y^2); \quad y = k/10 \hompc.
\] 
Thus the power needs to be boosted by about a factor 2 on the very smallest scales of all;
on the scales of interest here, such corrections to HALOFIT have a negligible impact.

\subsection{Robustness of modelling}
\label{sec:robustness}

Beyond any uncertainties in fundamental cosmological parameters, the dominant
potential source of imprecision in our lensing predictions comes from the 
imperfect knowledge of the true redshift distributions associated
with each tomographic slice, $N(z)$. Indeed, calibration of
the true redshift distribution associated with a
particular photometric-redshift selection is arguably the
dominant systematic in studies of weak gravitational lensing
(e.g. \citejap{KiDS450}; \citejap{DESy1lens}). This is why we
considered a number of different models for $dN/dz$ in
Section \ref{sec:dndz}, as shown in Fig. \ref{fig:dNdz_bins}.
The theoretical predictions were calculated assuming the 
different options from that section, asking whether the
changes in the predictions were significant in the context
of the statistical errors. In the interests of space we
will not present multiple versions of 
Fig. \ref{fig:corr_theory}. A brief summary of the 
findings is that the alterations in the harmonic correlation were
at the few per cent level for the different models, with the
exception of the simplest SDSS model based on the quoted
Gaussian error from \cite{Beck16}, where the changes with
respect to the direct GAMA-based calibration was at the 10\% level.
As will be seen below, the statistical uncertainties in the
amplitude of clustering, $\sigma_8(z)$ are at the 5--10\% level,
and we are therefore confident that remaining uncertainty
in photo-$z$ calibration is not important at the current level of
precision.

\begin{figure*}
\centering
\includegraphics[width=0.7\textwidth]{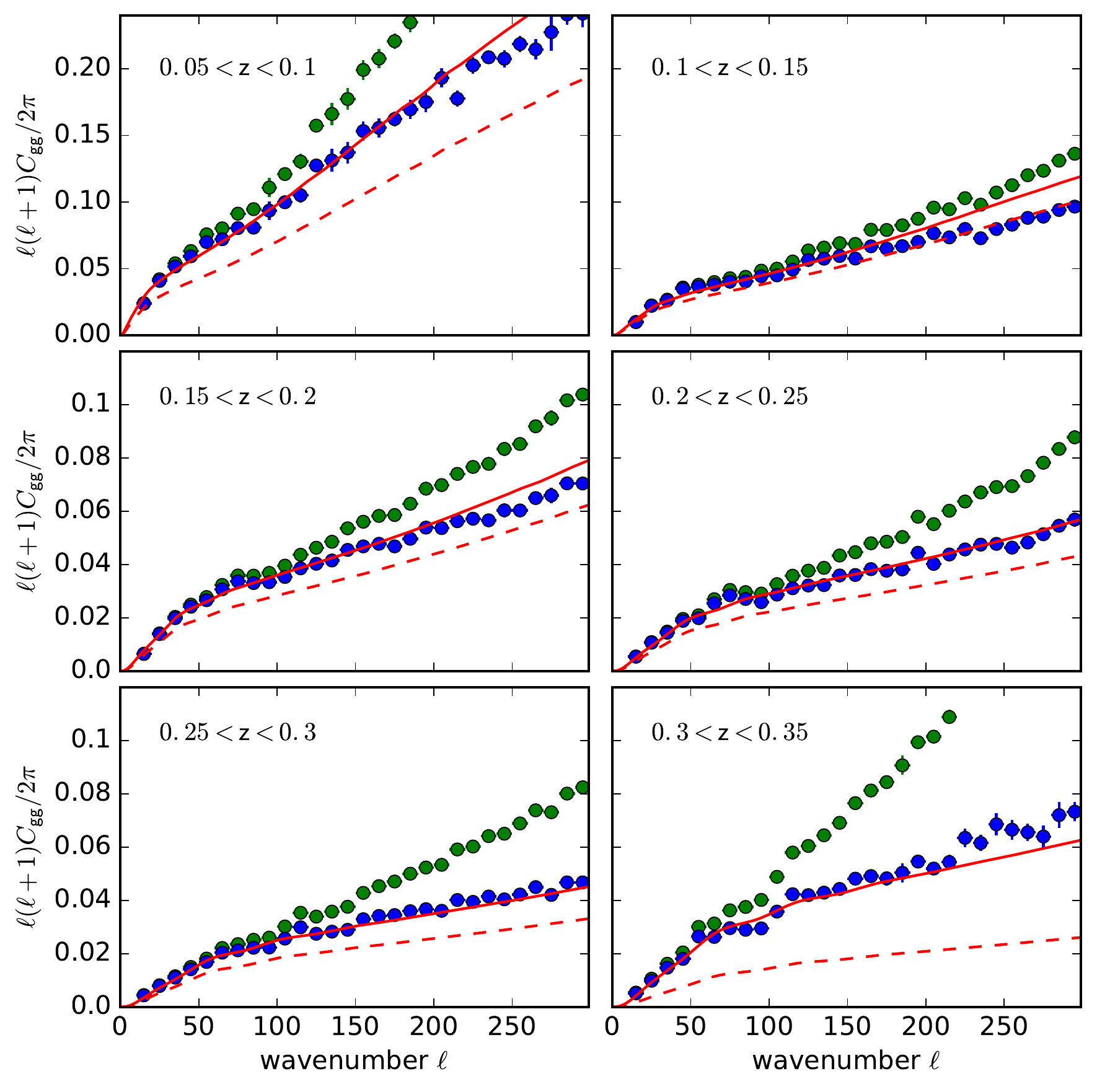}
\caption{The angular pseudo-spectra of the various 2MPZ ($0.05<z<0.1$) and \WIxSC\ tomographic slices.
The points show results without (green) and with (blue) shot-noise subtraction.
The red dashed line shows the predicted angular spectrum of the nonlinear mass
distribution, subject to the same $N(z)$ selection; the solid line shows the same
curve linearly biased to match the data at $\ell<150$. }
\label{fig:wisc_gg}
\end{figure*}

\begin{figure*}
\centering
\includegraphics[width=0.7\textwidth]{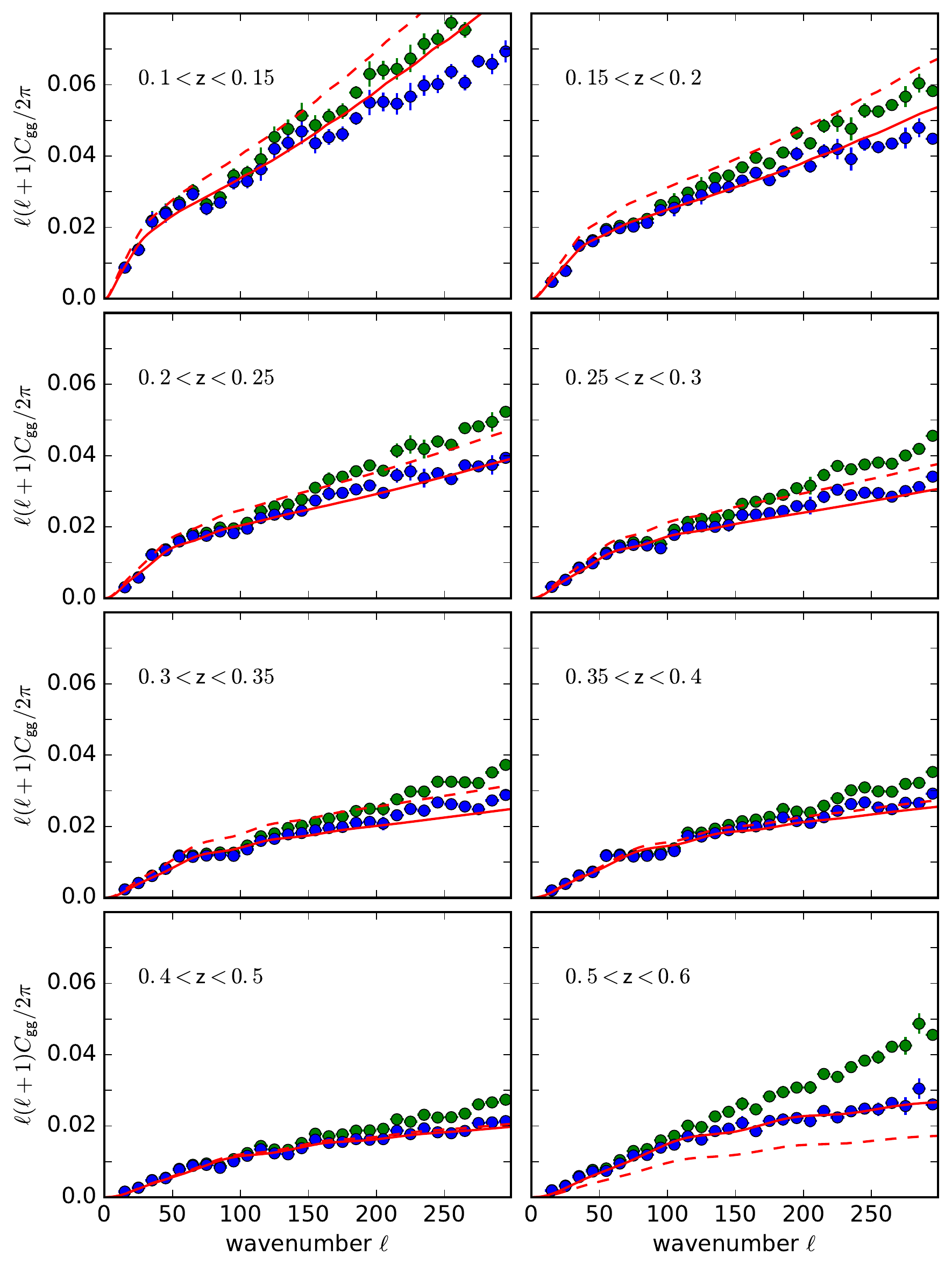}
\caption{The angular pseudo-spectra of the various SDSS tomographic slices.
The different points and lines have the same meaning as in the
\WIxSC\ data of the previous figure.}
\label{fig:sdss_gg}
\end{figure*}

\begin{figure*}
\centering
\includegraphics[width=0.7\textwidth]{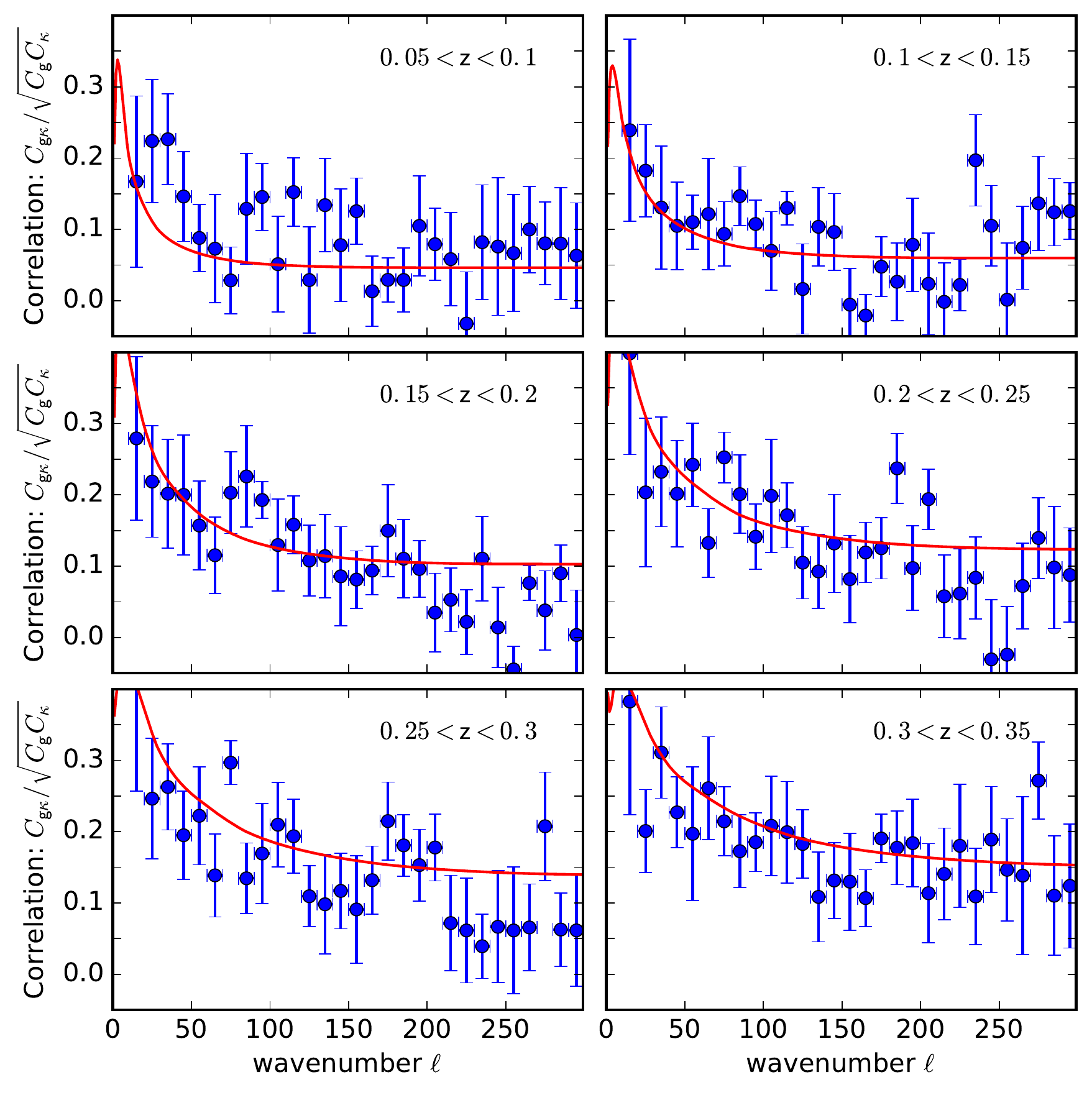}
\caption{The harmonic-space correlation coefficient
  $C_{g\kappa}\,/\,(C_{gg}C_{\kappa\kappa})^{1/2}$ for the various 
  2MPZ+\WIxSC\ tomographic slices.  $C_{g\kappa}$ and $C_{gg}$ are the direct
  pseudo-power estimates, and 
  we adopt the $\Lambda$CDM
  theoretical prediction for $C_{\kappa\kappa}$. 
  This statistic is independent of any degree of
  scale-dependent bias, $b(\ell)$, provided that it is not stochastic.
A correlation is detected with high significance in all slices, for all multipoles. 
The lines show the theoretical prediction of our fiducial $\Lambda$CDM
model, as presented in Fig. \ref{fig:corr_theory}, taking into account the
exact true redshift distributions for these slices.
}
\label{fig:wisc_corr}
\end{figure*}

\begin{figure*}
\centering
\includegraphics[width=0.7\textwidth]{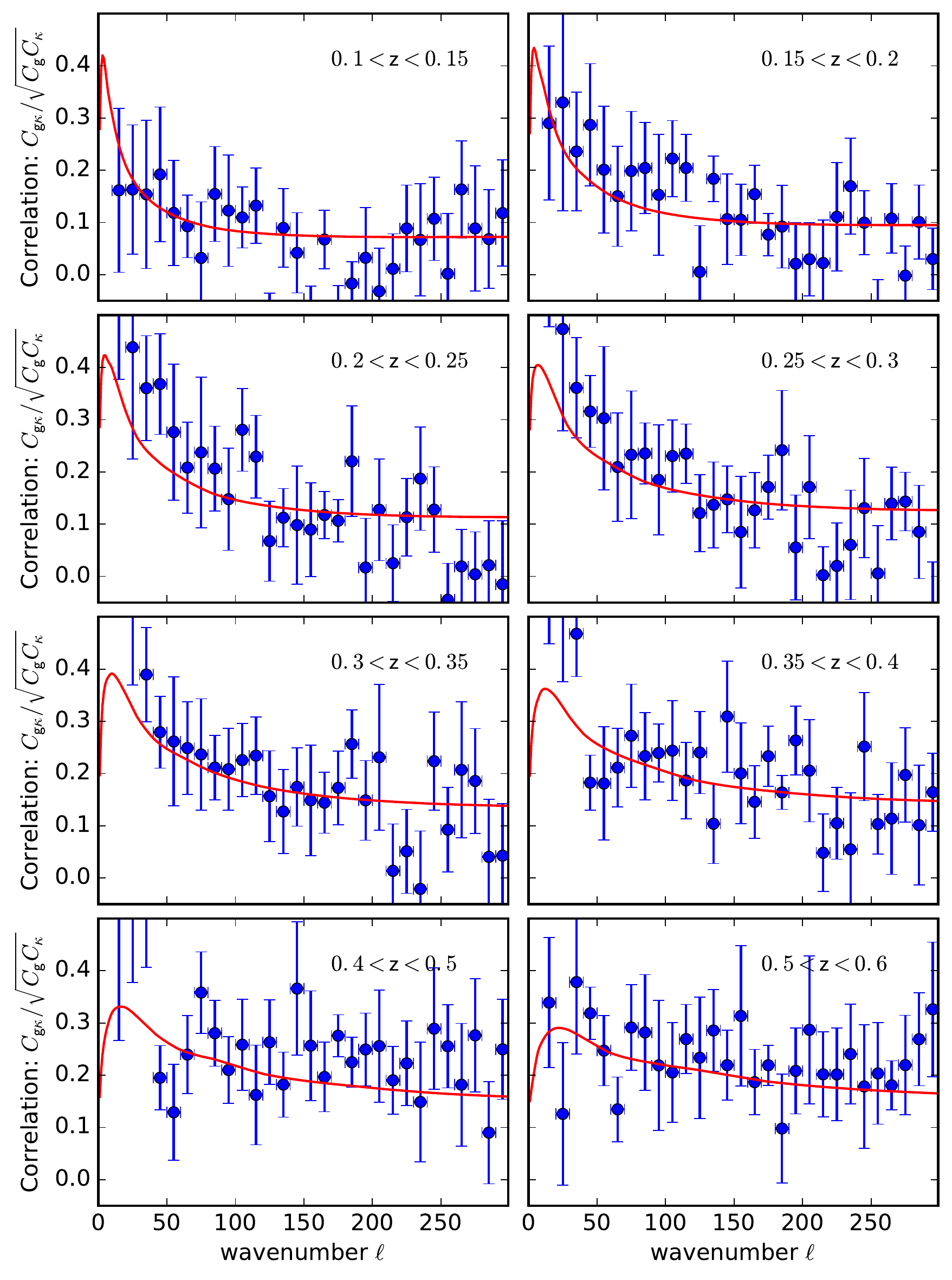}
\caption{The harmonic-space correlation coefficient
$C_{g\kappa}\,/\,(C_{gg}C_{\kappa\kappa})^{1/2}$ for the various SDSS tomographic slices.
The points and lines have the same meaning as in the
\WIxSC\ data of the previous figure.
}
\label{fig:sdss_corr}
\end{figure*}

\begin{figure*}
\centering
\includegraphics[width=0.65\textwidth]{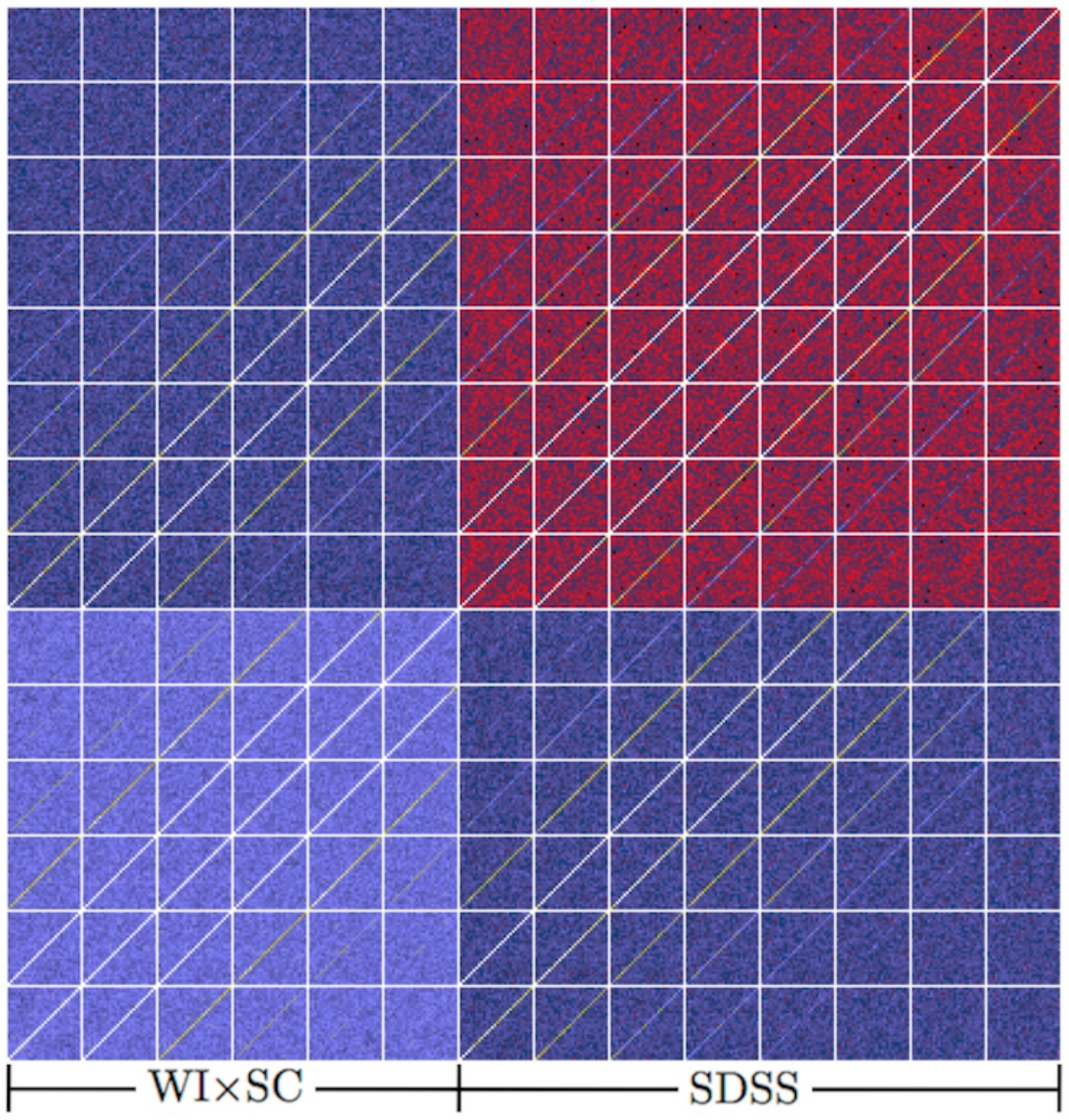}
\caption{The correlation matrix for the CMB lensing cross-correlation
  measurements.  Each square corresponds to a different pair of
  tomographic bins: the 2MPZ+\WIxSC\ bins constitute the lower left quadrant;
  the SDSS bins constitute the upper right. Cells increase with redshift
from left to right throughout a block. The colour coding is in part to
distinguish each block, rather than quantitatively encoding the correlation.
As expected, we see that
  adjacent bins are correlated (even extending over several bin
  separations), and that the 2MPZ+\WIxSC\ and SDSS results are correlated
  where their redshift bins are at similar distances. The typical
  correlation coefficients are about 0.7 at one bin separation and 0.5
  at two bins separation (similar in 2MPZ+\WIxSC\ and SDSS) and about 0.7
  between \WIxSC\ and SDSS in the same redshift bin. 
}
\label{fig:corrmat}
\end{figure*}

\begin{figure}
\centering
\includegraphics[width=0.48\textwidth]{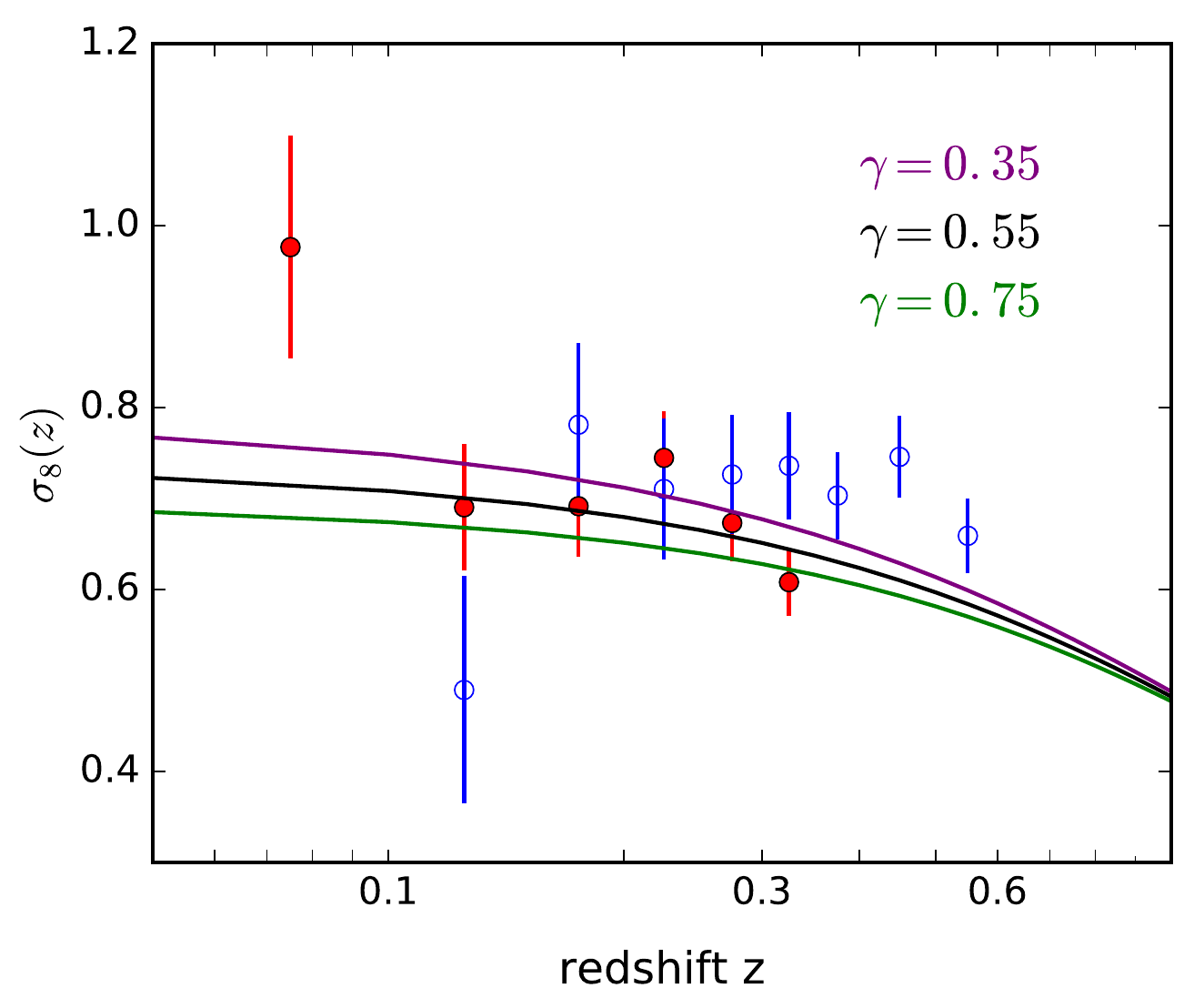}
\caption{A plot of the fluctuation amplitude as a function of redshift,
$\sigma_8(z)$, inferred from the amplitude of the cross-correlation results
shown in Figs \ref{fig:wisc_corr} \& \ref{fig:sdss_corr}.
2MPZ and \WIxSC\ results are shown as red solid symbols; SDSS as open blue symbols. These results
are derived by using the scalings from the fiducial model with assumed $\sigma_8=0.80$.
Solid lines show expectations for different values of $gamma$ (lower $\gamma$ yields
higher $\sigma_8(z)$).
}
\label{fig:evol}
\end{figure}

\section{Tomographic power measurements}

We now need to construct auto- and cross-power estimates from our various
tomographic slices to compare with the above models. This would
be straightforward if we had complete sky coverage, as we would
just construct the spherical-harmonic coefficients of the
observable quantity under consideration, $Q$:
\[ 
a_\ell^m=\int Q(\theta,\phi)\; Y^*_{\ell m}(\theta\phi)\; d\Omega,
\] 
where $(\theta,\phi)$ are polar angles, $Y_{\ell m}$ is a spherical
harmonic, and $d\Omega$ is an element of solid angle. Here,
$Q$ can be $\kappa$ or $\delta$, the surface density fluctuation
in a given tomographic slice. We would then construct a direct
estimator of the cross-power spectrum by averaging over $m$:
\[
{\hat C}_{\kappa\delta} = (2\ell+1)^{-1} \sum_{m=-\ell}^{m=\ell} a_\ell^m(\kappa)\, a^*{}_\ell^m(\delta)
\]
(and similarly for the auto-power spectra). This would normally
be presented as the power per $\ln\ell$:
\[
\hat{\cal P}_{\kappa\delta} = {\ell(\ell+1)\over 2\pi} \, {\hat C}_{\kappa\delta},
\label{eq:logpower}
\]
and this estimator would be suitable for direct comparison with the 
theory presented earlier.

The problem with this approach is that we have a mask that sets the
observable to zero over some region of the sky. Thinking in
Fourier language, this multiplication becomes a convolution
in harmonic space, and so there is a mixing: a single
$a_\ell^m$ coefficient for the direct transform of the masked data
is a linear combination of the coefficients for the full sky
(see section IV of \citejap{Peebles1973}).
But the impact of this mixing can be seen quite simply at
least in the limit of modes whose effective wavelength, $2\pi/\ell$,
is small compared to the scale of the mask. In equation 
(\ref{eq:logpower}), the factor $1/2\pi$ derives from 
the density of states -- i.e. we are simply saying that
the total power is the sum of the power from all the modes
in a given range of $\ell$. If a fraction $f_{\rm sky}$ of
the sky is masked, then the number of modes is reduced in
proportion to the sky area. Hence, the power obtained
from transforming the masked data is underestimated by
a factor $f_{\rm sky}$. We can therefore restore the
correct level of power by rescaling to obtain the
pseudo-$C_\ell$ estimator:
\[
{\hat C}_{\kappa\delta}({\rm pseudo-}C_\ell) = {1\over f_{\rm sky}}\,
{\hat C}_{\kappa\delta}({\rm masked})
\]
(see \citejap{Hivon2002}).
We neglect the very large-scale modes where this approximation
is less accurate, $\ell<10$, which are in any case very noisy. Elsewhere,
this simple approximation is adequate to much better than the precision of
the measurements.

This scaling also has implications for the errors on the measurements.
For Gaussian fluctuations (a reasonable approximation on most scales),
the fractional power errors in each $\ell$ bin are independent and
of amplitude simply $1/\sqrt{N_{\rm modes}}$, where there are
$N_{\rm modes}$ modes in the bin. For masked data, these
errors are therefore increased by a factor $1/\sqrt{f_{\rm sky}}$, 
assuming that the bin is wide in $\ell$ compared to the wavenumbers
on which the transform of the mask is significant.
In practice, we generate a covariance matrix so that the
correlation in errors between datasets can be assessed; but
for a single dataset,
this mode-counting argument works very well.

\begin{table}
\caption{Inferred values of bias as a function of redshift in the various
tomographic slices. We quote a measuring error, but it should be
noted that the true error is larger, as these bias values are with
respect to the fiducial model, with assumed $\sigma_8=0.80$. 
}
\begin{center}
\begin{tabular}{ccc} \hline
Dataset & $\langle z \rangle$ & $b(z)$
\\ \hline
2MPZ & 0.075 & $1.182\pm 0.009$ \\ \hline
\WIxSC & 0.125 & $1.086\pm 0.007$ \\
\WIxSC & 0.175 & $1.126\pm 0.007$ \\
\WIxSC & 0.225 & $1.144\pm 0.013$ \\
\WIxSC & 0.275 & $1.206\pm 0.009$ \\
\WIxSC & 0.325 & $1.548\pm 0.018$ \\ \hline
SDSS & 0.125 & $0.915\pm 0.010$ \\
SDSS & 0.175 & $0.894\pm 0.006$ \\
SDSS & 0.225 & $0.909\pm 0.007$ \\
SDSS & 0.275 & $0.902\pm 0.009$ \\
SDSS & 0.325 & $0.888\pm 0.013$ \\
SDSS & 0.375 & $0.966\pm 0.020$ \\
SDSS & 0.450 & $0.980\pm 0.019$ \\
SDSS & 0.550 & $1.245\pm 0.011$ \\
\hline
\label{tab:bias}
\end{tabular}
\end{center}
\end{table}

We now present the measured pseudo-power spectra in our various
tomographic slices. Figs \ref{fig:wisc_gg} \& \ref{fig:sdss_gg} show
the galaxy angular auto-power, $C_{gg}$, both the raw measurements
and corrected for shot noise: $C_{\rm shot}=4\pi f_{\rm sky}/N_g$,
where $N_g$ is the total number of galaxies in a given slice.
These galaxy spectra are contrasted with biased
non-linear mass correlations, showing that there is a weak
but significant scale-dependence of bias.
The degree of bias varies with redshift, tending to be anti-biased
for the low-$z$ slices and moving to $b>1$ at the higher redshifts.
This behaviour is reasonable: the combination of the flux limit
and the redshift limits means that the more local slices must consist entirely
of low luminosity galaxies, whereas the more distant slices can contain
some more luminous galaxies with large bias. For completeness,
we quote in Table \ref{tab:bias} the bias values inferred on the 
assumption of the fiducial model (fitting over $50<\ell<150$,
where there is no obvious scale dependence). However, we do not use these values directly. 

The existence of an a priori unknown degree of bias can be removed by
constructing a harmonic-space correlation coefficient 
similar to \eqref{eq:rmk} but using 
galaxy auto-power, $C_{gg}$, and galaxy-lensing cross-power,
$C_{g\kappa}$, instead of mass-based statistics:
\[
r_{g\kappa} = C_{g\kappa}\,/\,(C_{gg}C_{\kappa\kappa})^{1/2},
\]
and we present the results in this form
in Figs \ref{fig:wisc_corr} \& \ref{fig:sdss_corr}, rather than plotting
the raw cross-power. 
Here, $C_{g\kappa}$ and
$C_{gg}$ are the direct pseudo-power estimates. Although dividing by a
noisy quantity is in principle undesirable, the signal-to-noise of the
auto-power is very much higher than that of the cross-power, so that the
auto-power measurements effectively have negligibly small random errors.  The
same is very much not the case for the lensing auto-power, 
where the cosmological signal is at best comparable to the measuring
noise at $\ell \simeq 30$, and about 10 times smaller by $\ell=300$.
The raw measured power spectrum of the lensing map is thus biased well
above the level of the cosmological signal by the addition of the
noise auto-power, and using this raw power would inevitably
yield a small harmonic-space correlation -- in the same way
that the lensing and galaxy maps would show little correlation
in configuration space, because the former is noise-dominated.
We therefore choose instead to normalise our harmonic-space
cross-correlation by using
the $\Lambda$CDM theoretical prediction for
$C_{\kappa\kappa}$. 
Any uncertainty in this prediction is small compared to the
random errors in the cross-power, which dominate the uncertainty in
the correlation measurement.

This statistic has the virtue that it is independent of any degree of
scale-dependent bias, $b(\ell)$, provided that it is not stochastic.
In detail, we would not expect this to be the case: on scales where
nonlinearities are important, higher-order correlations affect the
galaxy-galaxy and galaxy-matter correlations differently, so that we
would not expect the identical $b$ to appear in
$C_{g\kappa}=bC_{\kappa m}$ and $C_{gg}= b^2C_{mm}$. But
on the scales and level of precision at which we are working, the
difference is negligible (\citejap{Modi17}), and so we can treat the
correlation statistic $r_{g\kappa}^2$ as measuring empirically the fraction of the
variance in $\kappa$ that is contributed by the tomographic slice
being studied. Note that the immunity to scale-dependent bias is
an advantage of this approach compared to a more formal method in
which the data for $C_{g\kappa}$ etc. are fitted directly, with bias
treated as a nuisance parameter to be marginalized over:
in that case, it is necessary to assume
that the bias is independent of scale (e.g. \citejap{DES_3x2pt2017}).

\subsection{Robustness checks}

\subsubsection{\WIxSC\ -- SDSS comparison}

The \WIxSC\ data yield good detections of the galaxy-lensing cross-correlation
in all tomographic bins, with a precision that is greater than the SDSS
results at the same redshifts, as expected from the greater sky coverage.
Given that both the legacy photographic optical data
and the low-resolution WISE measurements have their issues, especially
in terms of stellar contamination as discussed above, it seemed prudent
to check that these measurements are free of significant systematics.
We approached this by using SDSS data to create an ideal \WIxSC\ dataset
within the SDSS area. Taking the known colour equations \citep{Peacock16}, SDSS data were
used to generate SuperCOSMOS $B$ and $R$ magnitudes, which were then
degraded to match the measuring errors of the original photometry,
as quantified by \cite{Peacock16}. This catalogue was then extinction-corrected
and cut to the SuperCOSMOS limits, following which it was paired with WISE.
Finally, photometric redshifts
were estimated using the same ANNz code as for the real \WIxSC\ catalogue.
The power spectra for this idealized catalogue were then computed
and compared with the \WIxSC\ results, when restricted to the same sky
coverage as SDSS. The results for the $C_{g\kappa}$ and the
harmonic-space correlation were found to be in agreement to within
a small fraction of the measuring error. This test is not perfect,
since it is dominated by the sky region where the SuperCOSMOS data
were best calibrated. The calibration was performed
with DR6, but the DR12 release did not greatly extend the
area of the imaging dataset (9376 deg$^2$ for the BOSS area,
as opposed to 8417 deg$^2$ of legacy imaging in DR6).
However, as described in \cite{Peacock16}, the calibration in
the remainder of the sky was constrained primarily by optical--2MASS
colours, and the reliability of this strategy could be validated
using the plates with direct SDSS calibration.
There should thus be no concern about the photometric calibration
outside the areas with SDSS overlap. The direct \WIxSC--SDSS
comparison is useful because it addressed the impact of other
factors: poorer depth and poorer star-galaxy discrimination
in the legacy photographic data. But the results of this
section show that these factors do not have a significant impact
on the cross-correlation statistics studied here.
We therefore see no reason why the \WIxSC\ and SDSS data should not
be treated as a consistent whole, with the superior depth of
SDSS allowing our tomographic shells to be extended to higher
redshift over a smaller area.

\subsubsection{Thermal Sunyaev--Zel'dovich contamination}

A distinct possible concern is that the measured tomographic
cross-correlations may not reflect purely the desired
cross-correlation of density and lensing convergence. This is because
the CMB lensing map is constructed from non-Gaussian signatures in the
temperature and polarization maps, and there are other possible
non-Gaussian contributions beyond lensing. The {\it Planck\/} reconstruction
masks out known point sources, but this process may be incomplete; in
particular, there may be a contribution to the lensing reconstruction
from the thermal Sunyaev--Zel'dovich \citep[tSZ,][]{SZ} signal, which will also correlate
with the galaxy density. The extent of such leakage was considered by
\cite{Geach2017}, who estimated that it should bias the reconstructed
$\kappa$ by only a few per cent at the location of clusters, which
would be insignificant at the $S/N$ of our measurements.
This effect was considered in more detail by
\cite{MadHill2018}, who claimed that the magnitude of the leakage
increased for more massive haloes. We therefore decided to check the
impact of this effect on our results empirically, as follows. We identified the 1\%
highest density pixels in our tomographic slices and masked them out,
before repeating the cross-correlation analysis. This `censoring'
lowers the amplitude of galaxy clustering very substantially: a
reduction in linear bias by about a factor 1.5, and a reduction in
high-$\ell$ auto-power by over a factor 2. We have argued that our
harmonic-space correlation measure should be independent of such
scale-dependent bias effects, and indeed we found that the correlation
amplitudes from this modified analysis were unchanged in amplitude to
within negligible shifts of approximately 3\%.  This argues directly
that any tSZ leakage into the lensing map at high-density regions does
not cause a significant bias in our cross-correlation results.

This result can also be used to argue that other possible
biases associated with nonlinear regions are negligible.
The {\it Planck\/} lensing reconstruction works in the
limit of weak deflections, and so can in principle be biased
by neglected higher-order corrections. These have been estimated
to alter the inferred lensing power spectrum by a few \%,
which would be unimportant with current data 
\citep{Beck2018, Bohm2018}. However, \cite{Bohm2018} also suggest
that the effect of non-Gaussianity might be larger for 
low-$z$ cross-correlation studies, which is a possible concern.
\cite{Bohm2018} leave a detailed calculation of this effect for future work,
so at present we can only note that a bias associated with non-Gaussian
structures would also presumably reveal itself in the most
high-density regions. Thus the fact that we see no corrections
when removing such regions argues that any such effect is presently
unimportant. But all such biases will need to be considered
more carefully with future improved lensing data.

\section{Model fitting}

The visual impression of Figs \ref{fig:wisc_corr} \&
\ref{fig:sdss_corr} is of good agreement between the measurements and
our fiducial $\Omega_m=0.3$ $\Lambda$CDM model, but we now need to
quantify this; thus a covariance matrix for the various measurements
is required. In total there are 406 points to consider, as we have 14
tomographic bins and use 29 angular bins in each ($\Delta\ell=10$ up
to a maximum $\ell=300$, but omitting the lowest bin where the
pseudo-power estimate may be biased by the limited sky coverage).  The
most robust way to determine the covariance is by averaging over many
realizations of mock data, and this is relatively easy in this
case. The $S/N$ of the cross-power measurements is very much lower
than the other ingredients, so we simply make Gaussian realizations of
fake random lensing skies using the known total $S+N$ lensing power
and correlate these with the observed galaxy data, ignoring the cosmic
variance in the latter. For future work of this sort, where the
lensing map is less noisy, one may want to make full realizations
including properly correlated mock galaxy slices and adding their
lensing signal to the CMB realizations
\citep[e.g.][]{FLASK}. But this level of detail
would be overkill for the present application.
Our simple procedure generates a covariance matrix
for all the tomographic slices, shown in
Fig. \ref{fig:corrmat} as the correlation matrix, $C_{ij}/[C_{ii}C_{jj}]^{1/2}$. 
As expected from the large-area coverage, it
can be seen that adjacent $\ell$ bins are uncorrelated. However,
different tomographic slices are correlated via the tails of the
redshift distributions.

To be used in generating a likelihood $\propto\exp(-\chi^2/2)$, the
covariance matrix needs to be inverted.  Even with a large number of
data realizations, this inverse can be biased high and cause the
errors on fitted parameters to be underestimated
(\citejap{hartlap}). Alternatively, one can exploit the fact that the
covariance matrix is dominated by a set of diagonal lines where a
given $\ell$ bin is correlated only with bins of the same $\ell$ over
all the various slices. Setting the covariance to zero outside these
lines gives a very well-defined inverse, and the
agreement with the direct inverse is good, apart from verifying the
correction in amplitude predicted by \cite{hartlap}.
The resulting errors on the power in a given $\ell$ bin are close
to the naive estimates that one would make in the absence of a covariance
matrix: calculate the standard deviation in power over the modes in the bin,
divide by the square root of the number of independent modes, and divide
by the square root of $f_{\rm sky}$ to allow for the fact that the
number of effective independent modes is reduced according to the
area of sky covered.

But even though it is thus straightforward to compute the $\chi^2$ fit
between the fiducial model and the data in a single tomographic bin,
the full covariance matrix is essential in order to use 
all our data while allowing for the correlations induced by the overlap
in redshift distributions. This overall measure of fit is completely
satisfactory: $\chi^2=400$ for 406 degrees of freedom. We can also
ask if the separate parts of the data agree with the fiducial model,
computing $\chi^2$ for the 2MPZ+\WIxSC\ and SDSS components
separately. These figures are given in Table \ref{tab:par_fits}, and are both in
complete consistency with the fiducial model.

\begin{table}
\caption{$\chi^2$ values and fitted growth-rate parameters. 
These are derived from the cross-correlation data and models
shown in Figs \ref{fig:wisc_corr} \& \ref{fig:sdss_corr}, scaling the
models according to non-standard growth laws (compared to
$\gamma=0.55$ and a fiducial $\Omega_m=0.3$), and computing
$\chi^2$ using the covariance matrix from Fig. \ref{fig:corrmat}.
}
\begin{center}
\begin{tabular}{cccc} \hline
Dataset & $N_{\rm df}$ & $\chi^2_{\rm min}$ & parameter
\\ \hline
2MPZ+\WIxSC & 174 & 184.6 & $\gamma=0.79 \pm 0.19$ \\ 
SDSS & 232 & 206.5 & $\gamma=0.26\pm 0.21$ \\ 
All & 406 & 400.0 & $\gamma=0.77 \pm 0.18$ \\
\hline
\label{tab:par_fits}
\end{tabular}
\end{center}
\end{table}

\begin{table}
\caption{Inferred values of $\sigma_8(z)$, derived
from the amplitude of the cross-correlation results
shown in Figs \ref{fig:wisc_corr} \& \ref{fig:sdss_corr}.
These results are derived by
applying the observed ratio between our 
measurements and the fiducial model with assumed $\sigma_8=0.80$.
}
\begin{center}
\begin{tabular}{ccc} \hline
Dataset & $\langle z \rangle$ & $\sigma_8(z)$
\\ \hline
2MPZ & 0.075 & $0.977\pm 0.122$ \\ \hline
\WIxSC & 0.125 & $0.691\pm 0.070$ \\
\WIxSC & 0.175 & $0.692\pm 0.056$ \\
\WIxSC & 0.225 & $0.745\pm 0.052$ \\
\WIxSC & 0.275 & $0.673\pm 0.042$ \\
\WIxSC & 0.325 & $0.608\pm 0.037$ \\ \hline
SDSS & 0.125 & $0.507\pm 0.129$ \\
SDSS & 0.175 & $0.809\pm 0.092$ \\
SDSS & 0.225 & $0.735\pm 0.080$ \\
SDSS & 0.275 & $0.752\pm 0.068$ \\
SDSS & 0.325 & $0.762\pm 0.061$ \\
SDSS & 0.375 & $0.728\pm 0.050$ \\
SDSS & 0.450 & $0.772\pm 0.046$ \\
SDSS & 0.550 & $0.682\pm 0.043$ \\
\hline
\label{tab:sigma8}
\end{tabular}
\end{center}
\end{table}

\subsection{Implications for clustering evolution}

The results of the model fitting can be presented in a number of ways.
The simplest is to treat each tomographic slice independently and
measure the ratio between the cross-correlation signal and the
fiducial prediction. Multiplying this ratio by the fiducial
evolution of clustering yields an estimate of $\sigma_8(z)$ for that
bin; these values are collected in Table \ref{tab:sigma8}
and plotted in Fig. \ref{fig:evol}.
Visually, there is good overall agreement with the standard
$\gamma=0.55$ model.

\begin{figure}
\centering
\includegraphics[width=0.48\textwidth]{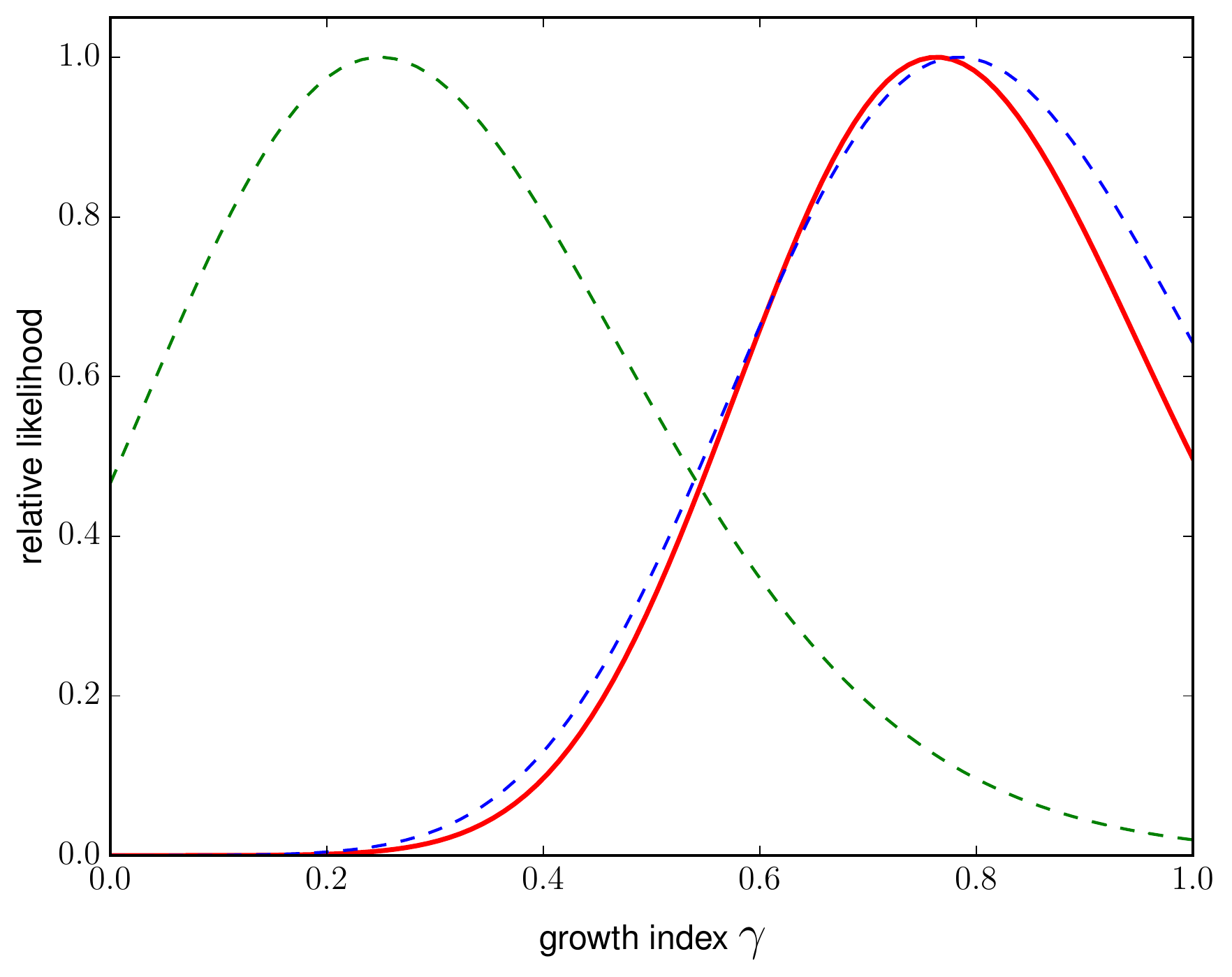}
\caption{Conditional relative likelihoods for $\gamma$ at fixed
  $\Omega_m=0.3$.
  Red solid lines show the result from the total
  sample; dashed lines show the separate results from \WIxSC\ (blue) and
  SDSS (green). 
}
\label{fig:condpars}
\end{figure}

The detailed conditional posterior for $\gamma$ at fixed
$\Omega_m=0.3$ is shown in
Fig. \ref{fig:condpars}.
The overall growth index from the combined \WIxSC\ and SDSS data is
slightly above the standard value: $\gamma=0.77\pm 0.18$, but a $1.2\sigma$
deviation is hardly to be regarded as surprising.
This value is deduced freezing the parameters of the fiducial cosmology,
as discussed earlier, although the precision on
$\gamma$ is sufficiently relaxed that any uncertainty on the 
fiducial model is unimportant.
The results for the \WIxSC\ and SDSS separately are consistent with $\gamma=0.55$ 
and with each other, although SDSS prefers a value below the standard one
whereas the value from \WIxSC\ alone is close to the overall value.
Indeed, the SDSS measurements alone are consistent with 
$\gamma=0$, so that a non-evolving fluctuation amplitude could not be
excluded using that subset of the data. At the current
level of precision, the adoption of the $\gamma$ model is thus
driven as much by external theoretical considerations as by
direct indications from the data. But the precision is
such that the SDSS and \WIxSC\ results are not in conflict, and the
overall measurements do strongly require evolution (non-zero $\gamma$).
This reflects the greater \WIxSC\ sky coverage, plus the fact that the
higher-$z$ SDSS data have less sensitivity to $\gamma$
(because $\Omega_m(z)$ approaches unity at higher $z$).

We have argued that these conclusions are not affected by the remaining uncertainties
in the fiducial cosmology, but the overall theoretical framework is still
critical. In particular, the $f_g=\Omega_m(z)^\gamma$ model implies
that  $\sigma_8(a)\propto a$ applies
for $z\gs1$, so that the expected amplitude of the cross-correlation
signal near the upper limit of our data is robustly predicted from the CMB. But if
we were to abandon the information in the absolute amplitude, and simply look
at the relative evolution of the cross-correlation signal with redshift,
then our constraints would be much less precise. This can be seen in
Fig. \ref{fig:evol}, where allowing 
an arbitrary vertical shift in the models would clearly remove much of
the sensitivity to $\gamma$.

\section{Summary and discussion}

We have presented a study of the cross-correlation between the {\it Planck\/}
map of CMB-inferred gravitational lensing and tomographic
photometric-redshift slices over $\simeq 70\%$ of the sky out to redshift
$z=0.4$ derived from the 2MASS \phz\ data, as well as the match between WISE and the SuperCOSMOS
all-sky galaxy catalogue, supplemented out to $z=0.6$ by SDSS 
\phz\ data over a smaller fraction of sky ($\simeq 25\%$).
We have carried out various investigations of the robustness of
these results, showing that the \WIxSC\ and SDSS datasets we use are consistent where they overlap;
that the true redshift distributions in our photometric slices are known to
sufficient precision; and that possible leakage of
Sunyaev--Zel'dovich signal into the cross-correlation measurements
is empirically negligible.

These results extend to higher redshift the similar
cross-correlation study between the CMB and the local
2MASS galaxy distribution using the 2MPZ dataset
(\citejap{2MxPlanck}), and they provide coverage of a much larger area
than the cross-correlation with the deeper DES data
(\citejap{Giannantonio2016}). In fact, the ability of DES to probe
to higher redshifts than the present study is of limited value.
The normal assumption is that any modifications of gravity
become important at low redshift (being in some way connected
with the onset of cosmic acceleration), as expressed
by assuming the growth-rate model $f_g=\Omega_m(z)^\gamma$;
thus data from high redshifts, where $\Omega_m(z)$ is close to
unity, lose all sensitivity to $\gamma$. 
Of course, it is still of interest to
measure the high-redshift evolution, since it is always
possible that we will have a major surprise and fail to validate
$\delta\propto a(t)$.
Nevertheless, the redshift range
covered here, out to $z\simeq 0.5$, is probably the sweet spot 
for studies that aim to measure $\gamma$ as a proxy for
testing gravity. 

From this work, our overall conclusion is that we detect the 
contribution of low-redshift shells to the CMB lensing signal,
and that the linear evolution of density over this range
requires a growth index $\gamma=0.77\pm 0.18$. This
measurement is consistent with the standard $\gamma=0.55$
expected in Einstein gravity, and this conclusion is
in agreement with a range of other studies
(e.g. \citejap{Simpson2013}; \citejap{BOSSMG2018}).
At present, the indirect measurements of the growth
rate from redshift-space distortions are more precise than
the result reported here: \cite{BOSSMG2018} quote
$\gamma= 0.566\pm 0.058$. An independent cross-check is
always valuable, of course, but it is interesting to
ask if CMB lensing tomography could match or exceed the
precision currently offered by RSD. Improvements can
happen in two ways: (1) increase the volume of the
tomographic shells, to suppress cosmic variance;
(2) improve the $S/N$ of the lensing map, which is
presently very far from being cosmic variance limited.
As regards the first option, there is little that 
can be done at $z<0.35$, because \WIxSC\ is
virtually full-sky. For the SDSS shells out to $z=0.6$,
in principle the area could be expanded by
a factor $\simeq 3$ (although current data from
Pan-STARRS and DES would not achieve this, and further
deep imaging in the far south would be needed). Even
so, reducing the current errors from the SDSS slices
by a factor around 1.5 would not be sufficient to pull
the error on $\gamma$ below 0.1 -- and as we have discussed,
there is limited information to be gained on $\gamma$
by pushing to higher redshifts. 

Therefore the major scope for improvement in these studies
lies with the CMB. The current {\it Planck\/} lensing map is
a tremendous achievement, but it is dominated by the
effects of small-scale detector noise in the temperature
and polarization maps that are used to make the reconstruction.
This can be seen clearly when we compare with results from
new ground-based CMB measurements. The South Pole Telescope
has produced a CMB lensing reconstruction over 2500 deg$^2$
(\citejap{SPTlens2017}); as their Figure 6 shows, the SPT data
alone are as accurate as {\it Planck\/} all-sky at $\ell=500$, so that 
all-sky data of SPT quality would yield an improvement in
power accuracy of about a factor 4 in this regime. 
Next-generation experiments such as CMB S4
\citep{CMBS4}  will continue this trend.
The improvement would be smaller at the wavenumbers of
$\ell\simeq 150$, which is where the present study
derives most of its signal, but we would then be able
to use a wider range of wavenumbers and gain from a
larger number of modes. Such a gain would come at the
price of needing greater care in the treatment of
nonlinearities and how these are altered
by baryonic effects ($\ell=500$ corresponds to
$k=1.7\hompc$ at $z=0.1$). But in principle there seems
no reason why CMB lensing tomography should not attain
errors of a few per cent in $\gamma$. The
competition from RSD will not stand still,
and next-generation RSD projects such as DESI may be expected
to push the errors on $\gamma$ to 1\% or better
(\citejap{DESI}). But it is clear that future
experiments will have ever greater concerns over
systematics as their formal statistical errors
shrink, and so we may expect CMB lensing tomography
to play an important future role in the robust testing of
Einstein gravity.

\vfill

\section*{Acknowledgements}

We thank Duncan Hanson and Antony Lewis for patiently responding to our
questions concerning the {\it Planck\/} CMB lensing data.
We are grateful to Jim Geach and Blake Sherwin for helpful comments on the manuscript.
JAP was supported by the European Research Council under grant number 670193.  MB was supported by the
Netherlands Organization for Scientific Research, NWO, through grant
number 614.001.451, and by the Polish National Science Centre under
contract UMO-2012/07/D/ST9/02785.

\bibliographystyle{mnras}
\bibliography{wixsc_cmblens}

\bsp	
\label{lastpage}
\end{document}